\documentclass[acmsmall]{acmart}
\setcopyright{none}
\renewcommand\footnotetextcopyrightpermission[1]{} 
\usepackage{soul} 
\usepackage{amsmath,amsfonts}
\usepackage{algorithmic}
\usepackage{graphicx}
\usepackage{textcomp}
\usepackage{xcolor}
\usepackage{xspace}
\usepackage[most]{tcolorbox} 
\usepackage{graphicx}   
\usepackage{subcaption} 
\usepackage{siunitx}
\usepackage{fvextra}
\usepackage{adjustbox}
\usepackage{enumitem}
\def\BibTeX{{\rm B\kern-.05em{\sc i\kern-.025em b}\kern-.08em
    T\kern-.1667em\lower.7ex\hbox{E}\kern-.125emX}}

\newboolean{showcomments}

\ifthenelse{\boolean{showcomments}}
{\newcommand{\nbc}[3]{
 {\colorbox{#3}{\bfseries\sffamily\scriptsize\textcolor{white}{#1}}}
 {\textcolor{#3}{\sf\small$\blacktriangleright$\textit{#2}$\blacktriangleleft$}}
 }
}
{\newcommand{\nbc}[3]{}
}

\newcommand{\smallsection}[1]{\noindent\textbf{#1}}

\newcommand{\tool}{\textsc{AlignMind}\xspace}

\definecolor{custom-gray}{cmyk}{0, 0, 0, 0.7, 1.00}
\newtcbtheorem[no counter]{Summary}{\hskip-0.97em}{enhanced,drop shadow={black!50!white},
  coltitle=white,
  top=0.15in,
  attach boxed title to top left=
  {xshift=1.5em,yshift=-\tcboxedtitleheight/2},
  boxed title style={size=small,colback=custom-gray}
}{summary}

\newcommand{\keyfinding}[1]{%
  \textbf{Key Findings:} #1
}

\newcommand{\implications}[1]{%
  \textbf{Implications:} #1
}

\tcbset{
  promptstyle/.style={
    breakable,
    colback=blue!5!white,  
    colframe=blue!75!black, 
    fonttitle=\bfseries,    
    fontupper=\small,
    boxrule=0.3mm,          
    arc=1mm,                
    left=1.5mm, right=1.5mm,    
    top=1.5mm, bottom=1.5mm,     
    before=\vspace{4mm} 
  }
}

\DefineVerbatimEnvironment{MyVerbatim}{Verbatim}{breaklines=true}

\newcommand{\rqi}{Can \tool output high-quality requirements and natural language workflows during the requirements clarification and refinement process?}
\newcommand{\rqii}{Is the output of \tool grounded on user conversations?}
\newcommand{\rqiii}{What are the operational costs associated with \tool?}

\newcommand{\llama}{Llama3.3-70b\xspace}
\newcommand{\gptmn}{gpt-4o-mini\xspace}
\newcommand{\gpt}{gpt-4o\xspace}

\begin{document}
\title{Towards Conversational Development Environments}
\subtitle{Using Theory-of-Mind and Multi-Agent Architectures for Requirements Refinement}

\setcopyright{acmcopyright}




\author{Keheliya Gallaba}
\orcid{0000-0002-5880-5114}
\email{keheliya.gallaba@huawei.com}
\affiliation{%
  \institution{Centre for Software Excellence, Huawei Canada}
  \state{ON}
  \country{Canada}}

\author{Ali Arabat}
\orcid{0009-0005-3024-6882}
\email{ali.arabat.1@ens.etsmtl.ca}
\affiliation{%
  \institution{École de Technologie Supérieure}
  \state{QC}
  \country{Canada}}

\author{Dayi Lin}
\orcid{0000-0002-4034-6650}
\email{dayi.lin@huawei.com}
\affiliation{%
  \institution{Centre for Software Excellence, Huawei Canada}
  \state{ON}
  \country{Canada}}

\author{Mohammed Sayagh}
\orcid{0000-0002-2724-0034}
\email{mohammed.sayagh@etsmtl.ca}
\affiliation{%
  \institution{École de Technologie Supérieure}
  \state{QC}
  \country{Canada}}

\author{Ahmed E. Hassan}
\orcid{0000-0001-7749-5513}
\email{ahmed@queensu.ca}
\affiliation{%
  \institution{Queen's University}
  \state{ON}
  \country{Canada}}
  

\begin{abstract}

Foundation Models (FMs) have shown remarkable capabilities in various natural language tasks. However, their ability to accurately capture stakeholder requirements remains a significant challenge for using FMs for software development. 
This paper introduces a novel approach that leverages an FM-powered multi-agent system called \tool to address this issue.
By having a cognitive architecture that enhances FMs with Theory-of-Mind capabilities, our approach considers 
mental states and perspectives of software makers. 
This allows our solution to iteratively clarify beliefs, desires, and intentions of stakeholders, translating these into a set of refined requirements and a corresponding actionable natural language workflow in the often-overlooked requirements refinement phase of software engineering, which is crucial after initial elicitation.
Through a multifaceted evaluation covering 150 diverse use cases, we demonstrate that our approach can accurately capture the intents and requirements of stakeholders, articulating them as both specifications and a step-by-step plan of action.
Furthermore, the lexical richness is eight times higher when the \tool is used for refining requirements compared to the baseline.
While there is an associated overhead in terms of time and resource usage, measured by the number of model calls and token usage, our findings suggest that the potential for significant improvements in the software development process justifies these investments. Our work lays the groundwork for future innovation in building intent-first development environments, where software makers can seamlessly collaborate with AIs to create software that truly meets their needs.

\end{abstract}

\ccsdesc[500]{Computing methodologies~Artificial intelligence}
\ccsdesc[500]{Software and its engineering~Designing software}
\ccsdesc[300]{Software and its engineering~Software development techniques}

\keywords{Requirements Engineering, Artificial Intelligence, Foundation Models, Intent Alignment}
\maketitle

\section{Introduction}

The advent of large-scale foundation models (FMs), such as large language models (LLMs), is reshaping the software development process. By leveraging training on source code repositories and textual artifacts from the software development process, these models can support software makers in various tasks such as code generation~\cite{wei2024magicoder, luo2023wizardcoder, starcoder2}, code documentation~\cite{Geng2024}, and test generation~\cite{Schfer2024, Lemieux2023}. Inherent ability of FMs to process and generate natural language makes them promising candidates for interpreting diverse stakeholder inputs, a cornerstone of Requirements Engineering~\cite{Nuseibeh2000}.
However, despite the apparent synergy and their broad capabilities, FMs have not yet streamlined all parts of software development, particularly the nuanced demands of refining requirements.

In fact, \textbf{requirements refinement} (i.e., elaborating on requirements, decomposing them into more manageable parts, resolving ambiguities and conflicts)~\cite{DeAngelis2018, Li2015, deJong} is a critical yet overlooked phase of software development in the AI era.
In practice, as stakeholders articulate expectations, developers must manage ambiguities, inconsistencies, and incompleteness~\cite{wu2024} in emergent requirements by engaging in dynamic dialogues with stakeholders. 
Complete and consistent requirements remain a principal driver of project success~\cite{Fernndez2016}, as inadequate or incomplete requirements can lead to costly rework, misaligned functionality, and overall project failures~\cite{Montgomery2022}. 
However, taking existing requirements, which might have been initially elicited in a raw, vague, or high-level form, and making them precise, detailed, complete, consistent, and understandable requires capabilities beyond brief, one-shot interactions.

Current FMs—despite their capacity to generate coherent, context-sensitive outputs—often rush to generate solutions. Wu and Fard~\cite{wu2024} found that in more than 60\% of the problem statements that required clarifications, FMs still generated code rather than asking clarifying questions crucial for effective refinement.
Bajpai et al.~\cite{Bajpai2024} also observed that FMs may prematurely attempt to
resolve tasks, even when the provided information is insufficient.
FMs' dialogue mechanisms, trained for conversational brevity, limit deeper investigations into ambiguous or incomplete requirements. This constraint can hamper the discovery of nuanced software needs and, in turn, compromise solution quality.
In fact, researchers have identified that there is often a disconnect between developers’ expectations and the responses that they receive for software engineering tasks because FMs are prompted with insufficient context, specifications, or clarity~\cite{ehsani2025}.


Clarified and detailed \textbf{requirements} directly inform design and implementation.
For example, the user may request ``a system that can quickly search through my documents''. However, there are many ambiguities in this requirement that need clarifying before starting implementation, such as the type of documents to be searched, e.g., text, images, or mixed; the search criteria to support, e.g., keywords, metadata, or full-text; and latency expectations, e.g., milliseconds, seconds or minutes.
On the other hand, \textbf{intentions}, while being abstract and high-level, guide the overall direction and decision-making.
The request of a user, ``I want to build a website displaying the latest tech news every day,'' could be motivated by the user's need to stay up-to-date with technology news. Identifying this user's intention helps to offer alternative solutions that suit the purpose, such as setting up a workflow to send a daily tech news digest to the user.
This fixation on a solution prematurely has long been recognized as a pitfall in requirements gathering~\cite{masteringreq} and underscores the need for intent alignment~\cite{EDamian2003}.

While some recent research efforts have begun investigating more robust conversational strategies for FMs, particularly in coding tasks~\cite{mu2024, wu2024}, these often offer incremental improvements for specific, well-defined interactions, such as code generation for programming competition questions. However, the complexities of refining requirements for real-world software projects, which demand nuanced understanding, iterative clarification, and extended engagement, remain largely unaddressed. A holistic approach that facilitates this kind of dialogue, mirroring effective human-to-human interactions in this domain, has yet to emerge.

Addressing this critical gap for real-world requirements refinement, this paper introduces a novel methodology to overcome the short and often unproductive discussion patterns exhibited by existing FM-based solutions. We propose an interactive framework designed to enable extended discourse, pinpoint ambiguities, and systematically address inconsistencies in requirement statements.
By combining theory-of-mind (ToM) capabilities~\cite{amirizaniani2024llms} with a multi-agent system, our approach ensures comprehensive coverage of stakeholder needs while preserving the efficiency gains associated with automated dialogue support.
Through this framework, we aim to bridge the gap between the promise of FMs for automated software development and the realities of prolonged, detail-oriented conversations that are essential for robust requirements clarification and refinement.
Our work lays the groundwork for more intuitive and effective development environments where AI collaborators can deeply understand and co-create software aligned with true stakeholder intentions.

Specifically, we have implemented these ToM and multi-agent approaches in a tool called \tool.
This tool begins by taking an initial user requirement as input and then, through a series of clarifying questions, deduces the user's intent and determines the final requirements.
The requirements are then translated into a natural language workflow consisting of step-by-step instructions. This workflow serves a dual purpose: first, as a crucial validation artifact for the refined requirements, ensuring they are complete and actionable; and second, as an explicit representation of the plan to fulfill those requirements. We evaluate this integrated approach through the following three research questions: 

\begin{description}
  
    

    \item [RQ1] \rqi
  
    \underline{\textit{Motivation:}} We aim to evaluate how effectively \tool can improve the output during the requirements clarification and refinement process. 
    The evaluation should take into account both the generated requirements and the natural language workflow proposed by \tool, as the workflow's quality is indicative of how well the requirements have been clarified and operationalized.
    We want to compare the improvement provided by \tool compared to the baseline, which is directly prompting one FM to act as a requirements refiner.

    \underline{\textit{Results:}} 
    We first conduct an evaluation using a panel of three FM-powered judges, based on five rubrics (i.e., assessment criteria) and 150 diverse scenarios in which the users want to refine requirements before building a software-based solution. 
    From this evaluation, we find that \tool has significantly higher output quality compared to the baseline. 
    Then, as an objective measure, we compute the \textit{requirement richness} quantifying the vocabulary variety using lexical richness~\cite{Torruella2013, kyle2019measuring} in the final set of requirements output by the baseline system and \tool. We find that the lexical richness is eight times higher when the \tool is used for refining requirements compared to the baseline.
    Furthermore, \tool enables longer multi-round conversations compared to the baseline.
    
  \item [RQ2] \rqii

  \underline{\textit{Motivation:}}  FM-powered systems are prone to hallucination~\cite{Ji2023}. In the case of requirements refinement, the system could hallucinate new requirements that the user did not ask for during the conversation.
  Therefore, it is critical to investigate to what extent the final set of requirements produced by \tool is affected by hallucination. 
  For this purpose, we investigate whether the generated requirements by \tool are grounded in content provided by the user, in contrast to generating content that is inconsistent with the user's conversations. 
  
  \underline{\textit{Results:}}  In \textit{abstractive summarization}~\cite{Gupta2019}, a summary and its source document are deemed factually consistent when the summary does not introduce any information that is not already present in the source~\cite{Kryscinski2019}. 
Similarly, in our context, we have to ensure that the final set of requirements does not include any new requirements that were not previously mentioned during the AI-human conversation. 
By assessing this consistency through a panel of FM-powered judges, we find that \tool shows no tendency to hallucinate. The requirements generated by \tool maintain a level of consistency with previous conversations comparable to the baseline. In fact, both methods achieved a perfect consistency score in most scenarios.


  \item [RQ3] \rqiii

  \underline{\textit{Motivation:}}  Before introducing an automated system such as \tool to improve the requirements clarification and refinement process, organizations need to consider the associated costs in practice. Therefore, we set out to investigate the operational costs of leveraging \tool for requirements refinement tasks in real-world scenarios. 
  
  \underline{\textit{Results:}}  While \tool has greatly enhanced the richness of requirements, our findings suggest that achieving this improvement comes with a cost overhead that must be taken into account.
We observe that \tool makes 10.6 times more API calls than the baseline. It uses 30 times the number of tokens in comparison to the baseline, based on the median. Balancing cost and performance is crucial when implementing an FM-powered requirements refinement system in real-world applications.
\end{description}

The remainder of this paper is organized as follows: 
We
summarize related work in Section~\ref{sec:related}.
Section~\ref{sec:solution} describes our approach for refining requirements using a multi-agent system with ToM capabilities. 
Section~\ref{sec:eval_results} presents our experimentation setup and results.
Section~\ref{sec:discussion} discusses the broader implications of our work.
Section~\ref{sec:threats} outlines the threats to validity.
Section~\ref{sec:conclusion} concludes the paper.

\section{Background and Related Work}
\label{sec:related}

In this section, we situate our work in the context of the literature on using language models for requirements engineering and leveraging the ToM capabilities of language models.

\subsection{Language Models for Requirements Engineering}


While several studies~\cite{Arora2024, White2024,Ronanki2024} explored the use of language models in requirements elicitation, some recent studies~\cite{Luitel2023, Santos2024, mu2024, ruan2024, Fazelnia2024} have tried to resolve requirements-related problems with the assistance of language models.
Luitel et al.~\cite{Luitel2023} used language models to find potential incompleteness in requirements. Santos et al.~\cite{Santos2024} investigated the suitability of using LLMs with in-context learning to check the requirements satisfiability given a system specification and associated domain knowledge. 
Fazelnia et al.~\cite{Fazelnia2024} proposed to integrate Satisfiability Modulo Theories (SMT) solvers with LLMs to detect conflicting software requirements.
ClarifyGPT~\cite{mu2024} added requirements clarification for LLM-based code generation, improving task performance that can be verified by generating solution candidates and test cases. Ruan et al.~\cite{ruan2024} demonstrated that using LLMs for developer intent extraction for automated program repair tasks is effective.
However, their approach infers intent from project structure and program behaviour instead of natural language utterances.
Arora et al.~\cite{Arora2024} explored the potential of LLMs in driving RE processes and conducted a preliminary feasibility evaluation of integrating LLMs into requirements elicitation.
Recent work~\cite{White2024, Ronanki2024} has investigated prompt patterns suitable for requirements elicitation.

Although there have been attempts to use language models in the requirement elicitation process, we observe a gap in using LLMs' conversational ability to clarify and refine requirements iteratively.
Furthermore, the feasibility of FMs should be explored in requirements refinement when implementing end-to-end software projects, beyond competitive coding tasks and method-level code completion.



\subsection{Theory-of-Mind in Language Models}

Theory-of-Mind (ToM) refers to the process of inferring a user’s intents, beliefs, and goals from their utterances~\cite{amirizaniani2024llms}.
With the advancements in LLMs, the research community is showing an increased interest in investigating their ToM capabilities to guide human-AI interaction (HAI) research~\cite{wang2022mutual}.
Wang et al.~\cite{wang2021towards} harnessed linguistic features extracted from conversations between students and a virtual teaching assistant to gain insights into students' perceptions. Their aim was to integrate the Mutual Theory of Mind concept into human-AI interactions.
Jung et al.~\cite{jung2024perceptions} proposed a new framework to improve LLM's ToM reasoning.
Their approach is designed to deduce the beliefs of the agents in an interaction by inferring others' perceptions and isolating the context perceived by others.
More recently, Shi et al.~\cite{Shi2025} showed AI systems can develop a sophisticated "theory of mind" by combining information from multiple modalities (e.g., language, vision, gestures) and reasoning about the mental states of multiple agents simultaneously.
Wilf et al.~\cite{Wilf2023ThinkTP} demonstrated \emph{perspective-taking}, i.e., placing oneself in another's position, as a promising direction for improving LLMs’ ToM capabilities.
Their work is inspired by \emph{Simulation Theory}, the prominent cognitive science perspective which argues that perspective-taking is the initial step to simulating another’s mental state.
Amirizaniani et al.~\cite{amirizaniani2024llms} assessed the abilities of LLMs to perceive and integrate human intentions and emotions into their ToM reasoning processes for open-ended questions by using posts from Reddit's ChangeMyView platform.
Fang et al.~\cite{fang2024inferact} explored the potential of using the ToM capabilities of LLMs to proactively identify possible errors before crucial actions are taken. This approach aims to ensure that LLM-based agents can be effectively deployed in critical environments.
Alongside these studies, we hypothesize that LLMs' inherent ToM capabilities can be enhanced and effectively leveraged specifically for intent alignment and requirement refinement in the software engineering domain.

Based on insights from these related publications, we see a promising opportunity to illustrate the effectiveness of using a multi-agent system equipped with Theory of Mind (ToM) capabilities for requirements clarification and refinement through multi-round conversations.

\section{Solution Design}
\label{sec:solution}

This section discusses the iterative prototyping approach that we adopted to 
develop an improved FM-powered requirements refinement system, followed by a detailed discussion of the design decisions we made after the iterative feedback collection process.
The goal is to build an FM-powered system where users, via natural language, can refine the requirements of a software solution that they wish to be built. 
The final output of this system will be a set of refined requirements for a given initial user query and a natural language workflow to achieve these requirements.


\subsection{Iterative Prototype Development}\label{res:pq}

We adopted an iterative prototyping approach to develop a conversational FM-powered solution for refining requirements. By providing a prototype for users to engage with, we were able to pinpoint any challenges they may face during its use and make improvements based on their feedback. To facilitate this process, we recruited six software engineering experts. These participants worked with a prototype of our tool to clarify the requirements for a specific task that needed refinement. Using their insights, we continually enhanced the prototype and identified the core concepts that will eventually guide the implementation of our solution.

Our iterative prototyping process consisted of the following activities:

\begin{enumerate}
    \item Implement an initial prototype of an FM-powered requirement refinement system.
    \item Conduct an interview with a participant where they will first interact with the prototype and then provide answers to a questionnaire.
    \item Incorporate new feedback from the participant into the prototype 
    \item Iterate from step 2 until all six participants are interviewed.
\end{enumerate}

Next, we detail each step of the process.

\begin{figure}[h]
    \includegraphics[width=\textwidth]{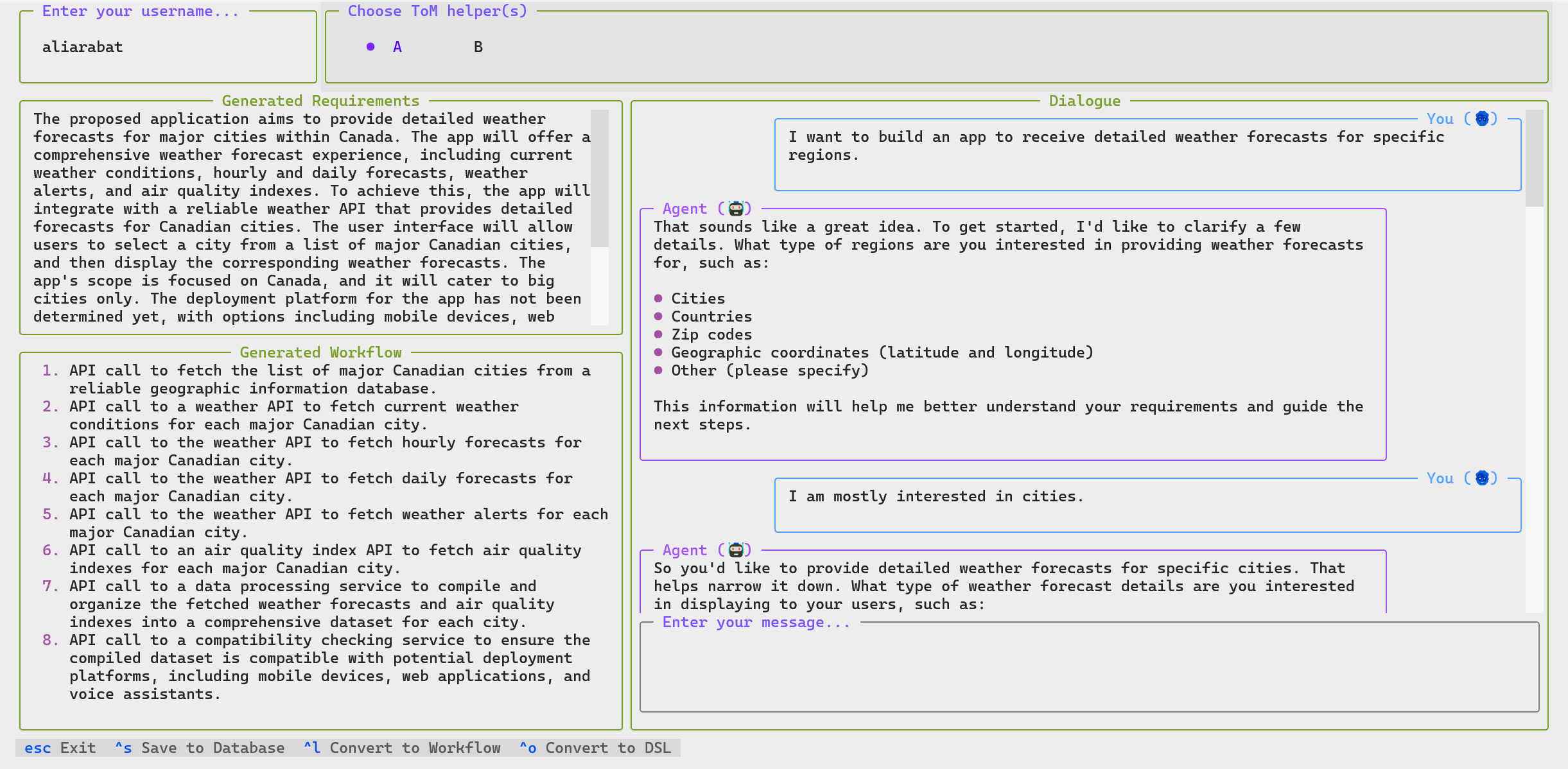}
    \caption{The Terminal-based User Interface (TUI) for \tool features a horizontal panel where users can input their username and choose configuration options. To the right, a chat box displays the conversation history with \tool, while a text field below allows users to type messages. On the left side, \tool displays the generated requirements and workflow.
    }
    \label{fig:tui}
\end{figure}

\noindent\textit{\underline{Initial Prototype Construction.}} To facilitate user feedback collection about FM-powered requirements refinement, we developed a Terminal-based User Interface (TUI), implemented using \textit{Textual}~\footnote{\url{https://textual.textualize.io/}} framework, as shown in Figure~\ref{fig:tui}. 
The TUI consists of three main panels. The first panel, positioned horizontally, allows the user to input their name and select a configuration. This configuration option helped us to present different backend implementations to the user and observe the interactions. The second panel on the right features an input field for users to express their intents in natural language (i.e., ask a question or seek clarification) and a pane displaying past conversation history between the user and the prototype. Finally, the `Requirements' and `Workflow' panels on the left are progressively updated as sufficient information about the user's intent is gathered. 

The initial version of the prototype directly forwarded the user's queries to an FM along with a system prompt, which requested the FM to gather users' requirements and respond with requests for clarification as needed or a finalized set of user requirements and a workflow in natural language that is necessary to achieve the requirements. The full system prompt is available in the Appendix~\ref{sec:baseline-prompt}.

\noindent\textit{\underline{Participant Recruitment.}} Via communication in internal mailing lists of a tech company, we recruited six individuals to converse with the requirements refinement system and complete the questionnaire. The software development experience of the participants ranges from 4 to 20 years, with a median of eight years. Their experience using LLMs spans from 5 months to 24 months, with a median of 14 months. 

\noindent\textit{\underline{User Interviews.}} The first two authors conducted face-to-face interviews with the participants. The interviews lasted between 60 and 90 minutes and were divided into two parts. The first part of the interview consisted of interaction with the prototype.
In this session, the participants were presented with an initial set of ten example tasks that required refinement and spanned multiple domains. The users were asked to choose one of these tasks or create one on their own, inspired by the examples provided.
Participants then engaged with the prototype, with their session data being saved into
an SQLite database upon exiting the application.
The time taken and the number of conversation rounds involved in completing the requirements refinement by each user were recorded.
At the end of the interaction with the prototype, participants were asked to complete a questionnaire designed according to the procedures of conducting controlled experiments~\cite{Ko2015-xn}. 
The questionnaire included the following questions:

\begin{itemize}
    \item \textbf{Programming experience:} How many years of software development experience do you have?
    \item \textbf{LLMs experience:} For how many months have you been using LLMs?
    \item \textbf{Conversation:} How would you rate the quality of your conversation with the tool based on the following attributes (ignoring the generated requirements and the workflow)?: (1) coherence, (2) identification of interests, and (3) sufficiently detailed responses
    
    Choose from the following options: \textit{Strongly Dissatisfied} / \textit{Dissatisfied} / \textit{Neutral} / \textit{Satisfied} / \textit{Strongly Satisfied}.
    \item \textbf{Requirements:} How satisfied are you with the generated requirements? 
    
    Choose one of the following options: \textit{Strongly Dissatisfied} / \textit{Dissatisfied} / \textit{Neutral} / \textit{Satisfied} / \textit{Strongly Satisfied}.
    \item \textbf{Workflow:} How satisfied are you with the generated workflow?

     Choose from the following options: \textit{Strongly Dissatisfied} / \textit{Dissatisfied} / \textit{Neutral} / \textit{Satisfied} / \textit{Strongly Satisfied}.
    \item \textbf{Strengths:} What did you like the most about the tool?
    \item \textbf{Weaknesses:} What challenges or issues did you encounter when using the tool?
    \item \textbf{Future use:} Would you consider using the tool to refine your requirements in the future?
    \item \textbf{Overall experience:} Can you briefly describe your overall experience with the tool in two or three sentences?
\end{itemize}

\noindent\textit{\underline{Post-user study analysis.}} After each participant had completed the questionnaires, we transcribed the responses from all participants into a structured document to draw further conclusions. Then, the authors discussed the transcribed responses to determine if any changes should be made to the prototype. Once the final prototype was obtained, we extracted the list of features of this prototype, then used open coding to group related features into more abstract concepts.


\noindent\textbf{\textit{Observation 1:}} \textbf{All participants agreed on the usefulness of using an FM-powered tool for requirements refinement.} In the responses, the participants appreciated the convenience offered by such a tool. 
For example, P3 stated, ``It does help you reflect on and reason about the requirements. Maybe the nicest aspect was the final list of requirements. I’m not sure if I would have been able to create such a detailed list myself (at least not in one go).''
The participants echoed the sentiment that such a tool helps the user by guiding them through the requirements clarification and refinement process, as shown by the comment by P5, \textit{``[tool] seems promising in assisting users to refine requirements for achieving a complex goal.''}

\noindent\textbf{\textit{Observation 2:}} \textbf{Participants identified several challenges with the requirements refinement system prototypes, such as lack of depth solutions, repetition, and unnatural conversations.}
Two participants found the requirements generated by the tool weren't as in-depth as they wanted. As P1 puts it, \textit{``[tool] ended up coming up with a workflow that on the surface seemed to make sense. However, [tool] seemed to just be coming up with `cookie-cutter' requirements.''}.
Another comment by the same participant, \textit{``[tool] seemed to prematurely arrive at a workflow''} may explain the lack of depth in the generated output.
Two of the participants found the tool repetitive at times. P4 explains this as follows: \textit{``[Tool] had more repetitive questions, sometimes felt like repeating myself multiple times.''}
P3 mentioned that the conversation \textit{``feels a bit too templated/scripted.''} The same participant explained a scenario in which the conversation flow with the tool broke when they tried to add additional details to a question, and the tool ended up responding to two things simultaneously. This observation emphasized the need to confirm with the user before moving to the next topic during a conversation.

\subsection{Four Key Pillars of our Solution}
\label{sec:concepts}

The final prototype that we obtain after the iterative prototyping phase is named \tool and is capable of systematically capturing requirements through dialogue and developing a refined workflow. 
We identify multiple distinguishing features of \tool compared to directly prompting FMs.
\begin{itemize}
     \item \textbf{Multi-Agent Architecture.} FMs often struggle with handling long, multi-objective prompts effectively~\cite{shiicml23}. Decomposing the requirements refining task into smaller, independent components improves performance. Therefore, we implement a multi-agent architecture where each agent has a distinct, well-defined role.
     \item \textbf{Theory-of-Mind.} Effective AI-human collaboration requires understanding the implicit human intent beyond what has been explicitly communicated. Prior research in Human-AI Interaction (HAI) supports the idea that AI systems with ToM capabilities foster more constructive and coherent communication~\cite{wang2022mutual}. Hence, our solution incorporates ToM units that infer human characteristics, needs, and goals in the background, leading to more context-aware interactions and better-aligned responses in the requirement refinement task.
     \item \textbf{Iterative Improvement.} Users rarely articulate perfect requirements in a single step; refining them over multiple rounds enhances precision and clarity~\cite{mu2024}. Thus, our solution supports multi-round conversations while maintaining a persistent internal state across invocations. This enables users to iteratively refine their requirements and build a well-structured natural language workflow.
     \item \textbf{Intent Decomposition.} Breaking down complex problems into smaller sub-problems is a fundamental principle in computational thinking~\cite{polya1945solve}, leading to more manageable and accurate solutions. Accordingly, our approach decomposes user intent into subtopics and generates targeted questions for each, ensuring a more structured and thorough understanding of the requirements, ultimately improving the final generated workflow.
\end{itemize}

\subsection{Multi-agent Architecture} 
\label{sec:goal-align-sol}

\begin{figure*}[tb!]
    \centering
    \includegraphics[width=\textwidth]{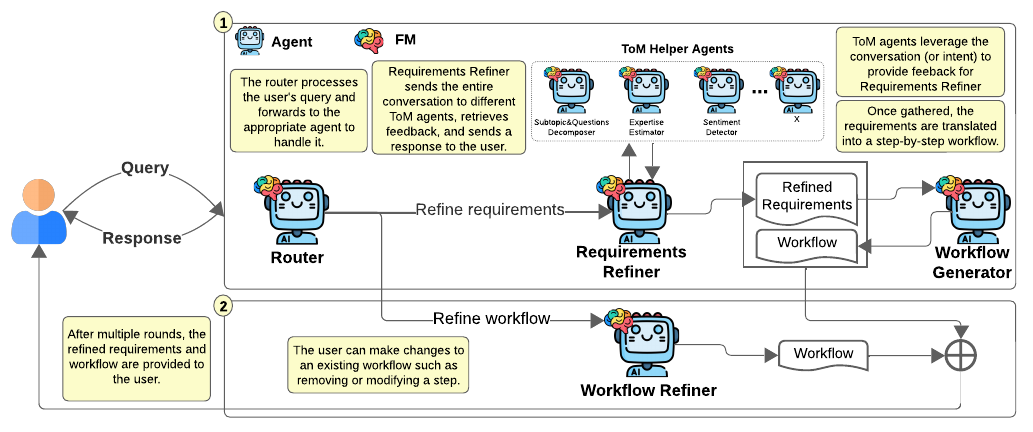}
    \caption{An overview of the proposed requirements refinement system, \tool, which consists of Router, Requirement Refiner, Workflow Generator, Workflow Refiner, and ToM helper agents.
    }
    \label{fig:ga_design}
\end{figure*}

During the process of refining user requirements, \tool has to achieve three objectives: maintaining a dialogue with the user while resolving ambiguities, preparing a complete summary of requirements, and generating a natural language workflow to achieve those requirements. However, we found that current FMs exhibit degraded performance when used as a single agent with very long, multi-objective prompts.
To address this, we propose an approach that involves multiple FM-powered agents, each responsible for a specific task. 
This approach aligns with the conventional Software Engineering wisdom of decoupling and helps ensure that changes to the system prompt won't interfere with the performance of another agent's capabilities.

To better explain our approach, we consider a running example as follows.
\begin{Summary}{Running Example}{}
A user submits the query, ``I want to build an app to receive detailed weather forecasts for specific regions.'' After multiple iterations with our multi-agent solution and internal processes, the user has refined requirements and a workflow with multiple steps.
\end{Summary}

Figure~\ref{fig:ga_design} illustrates different FM-powered agent components of the solution and the data flow across the agents of \tool when such a query is submitted.
Through multi-round interactions with the user, our solution refines the user's requirements and generates the final workflow to fulfill these requirements.
The main entry point to \tool is the Router Agent, which processes the user's query 
and directs it to the appropriate agent. If the query involves refining the user's requirements, it is handled by the \textit{Requirement Refiner Agent} and subsequently the Workflow Generator Agent; otherwise, the query is managed by the \textit{Workflow Refiner Agent}. To assist with the requirements refinement, the system also includes a group of ToM helper agents. We next explain each of the agents' behaviour in detail.

\subsubsection{Router Agent}
\label{sec:router}

The Router Agent receives a user's query and routes it to either the Requirement Refiner Agent or the Workflow Generator Agent based on the query's purpose. The system prompt associated with this agent defines a set of instructions for the FM to follow to identify whether a query should be sent to the Requirement Refiner Agent or the Workflow Refiner Agent. To help the FM make the right decision, we provide three few-shot examples, each of which illustrates the next agent to handle the query. The first example scenario is when a user asks for an update to the requirements, the second is when a user asks for a change to the workflow with no requirements, and the third is when a user requests to change the workflow when requirements have already been generated. To ease the post-processing of the Router Agent's response, we explicitly prompt the model to generate only the relevant agent to which we should forward the query (i.e., either ``RequirementRefiner'' or ``WorkflowRefiner''). For our running example, the output of the Router Agent would be ``RequirementRefiner'', indicating that the query should be forwarded to the Requirement Refiner Agent.


\subsubsection{Requirement Refiner Agent}
\label{sec:req_ref_agent}

The goal of the Requirement Refiner Agent is to interact with users through a set of clarification questions to clarify users' intent and develop a set of detailed requirements.
The refinement process of requirements is carried out iteratively, as shown in the top portion of Figure~\ref{fig:ga_design}. As the first step, the user interacts with the Requirement Refiner Agent to clarify their intent, supported by a set of FM-powered agents that are based on the ToM concept and further detailed in Section~\ref{sec:tom-solution}.
These agents infer different perspectives related to the context of the query.
Among such inferred perspectives are the user's interests, topics, goals, sentiment, and expertise.
The feedback from these ToM helpers is returned to the Requirement Refiner Agent, allowing it to provide a more accurate and targeted response to the user. The user can provide clarifications or ask back questions in natural language. Once the user’s intent is clarified through the dialogue, and when the Requirement Refiner Agent determines it has collected sufficient information, the Requirement Refiner Agent produces a detailed set of requirements, summarizing the dialogue and highlighting key points. This output is then passed to the Workflow Generator Agent.

\subsubsection{Workflow Generator Agent} Subsequently, the Workflow Generator Agent translates the requirements forwarded by the Requirements Refiner Agent into an actionable plan in natural language to achieve the user's requirement. This step is integral to the refinement process for three reasons: (1) It forces a check on the completeness and consistency of the requirements; if a coherent workflow cannot be generated, it indicates gaps in the requirements. (2) The generated workflow provides a more concrete artifact for stakeholder validation than textual requirements alone. (3) It represents the first step in operationalizing the user's intent, paving the way for future (semi-)automated execution. 
However, as observed in previous work~\cite{10.1145/3650105.3652301}, where the authors found that 26.4\% to 73.7\% of FM-generated output requires parsing or post-processing for code translation tasks, we observed that the Workflow Generator Agent can sometimes produce the workflow with additional text. Therefore, we implement a post-processing step that removes any free-form text and only retains the numbered steps as initially requested.

\subsubsection{Workflow Refiner Agent}
\label{sec:workflow_ref_agent}

A second use case of \tool involves the enhancement of an existing workflow. This flow is shown in the bottom portion of Figure~\ref{fig:ga_design}. For instance, a user may want to add missed steps during the initial requirements refinement, modify incorrectly defined steps or even remove unnecessary ones. The Workflow Refiner Agent handles these adjustments based on the user's prompts. For instance, a user might request: ``Can you change the third step in the workflow by replacing WeatherAPI with OpenWeatherMap API?''.
In such cases, the Workflow Refiner Agent processes the requested changes, adjusts the workflow, and returns an improved version in real-time.
This iterative cycle ensures that the user’s evolving preferences and requirements are continually incorporated into the workflow, maintaining alignment with their intent.

\subsection{Improvements based on Theory-of-Mind (ToM)}
\label{sec:tom-solution}

\begin{figure}[h]
    \includegraphics[trim={0.5cm 0.4cm 0.5cm 0.4cm},clip, width=\textwidth]{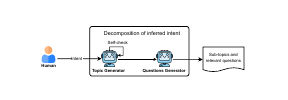}
    \caption{An overview of the topic and question generation workflow. 
    }
    \label{fig:tom_design}
\end{figure}

We use multiple ToM-based agent components to enhance the requirements clarification and refinement process. Accordingly, we provide a detailed definition of each component and its role in achieving the user's intent.


\subsubsection{Topics \& Questions Decomposer Agent} Upon the reception of the initial user requirement by the Requirement Refiner Agent, it forwards this initial requirement to the Topics \& Questions Decomposer Agent, which defines topics related to the user requirement and suggested clarification questions related to each topic. These topics and their related questions are to clarify the user intent to develop a clear and detailed requirement. This agent works in three steps: the agent generates three groups of topics, each group with a maximum of five topics. Then the agent self-reflects on the generated groups to identify an optimal group of a maximum of five topics. The optimal group is expected to cover a large spectrum of topics related to the user's initial requirement.
Then, we send these results to the question generator agent, which identifies a maximum of five questions for each of our five topics.
In our running example of the user asking to build a weather forecast application, the subtopics that would be generated are high-level, such as \textit{App User Needs and Goals}, \textit{Core Features}, \textit{Weather Data Sources and APIs}, \textit{Technology Stack}, and \textit{Deployment Platforms}. Each sub-topic is accompanied by questions for framing and clarifying the intent of the user. For the \textit{User Needs and Goals} topics, potential questions might include:

 \begin{itemize}
    \item What specific weather information does the user want (e.g., temperature, precipitation, wind speed, humidity)?
    \item What is the user's preferred frequency and granularity of forecasts (hourly, daily, weekly)?
    \item What are the user's desired regions? For example, are they local areas, global regions, or user-defined locations?
\end{itemize}

These subtopics and their corresponding questions are sent back to the Requirement Refiner Agent, which iteratively refines the requirements with the user across the subtopics independently.

\subsubsection{Users interaction with Requirement Refiner Agent for Clarifications:} While the Topics \& Decomposer Agent is invoked once to generate topics and their questions, the Requirement Refiner Agent sends back one question at a time to the user in an iterative fashion. This agent will also make sure that questions are not repeated, focus on one topic at a time, and track the evolution of the discussion with the user. During the conversation, \tool uses two strategies to determine if a certain subtopic is covered and that the conversation should move to a new subtopic. First, the Requirement Refiner Agent performs a self-check to determine whether the conversation history contains a sufficient number of question-answer pairs to cover each of the subtopics.
Second, \tool checks if a hard cut-off of $n$ questions has been spent inquiring about the sub-topic. We use $n=5$ in the experiments for this paper as a reasonable cutoff.
If either of the above two conditions is satisfied, the Requirement Refiner Agent moves the conversation to the next topic.

\subsubsection{Expertise ToM Helper:} After each user iteration, including the first one, the \tool sends the entire dialogue, including the current iteration, to the multiple ToM helpers for feedback, including the Expertise ToM Helper. The Expertise ToM Helper analyzes the user's language and conversational history, performing reasoning to classify the user's expertise level as one of the following: ``Novice'', ``Intermediate'', or ``Expert''.
This will eventually guide the Requirement Refiner Agent and Workflow Generator Agent to provide adequate responses that align with the expertise of the user. 
For the first iteration of our running example, the Expertise ToM Helper returns ``Novice''. 
Based on this estimated novice user experience, the Requirement Refiner Agent will ensure that the user is not overwhelmed with complex technical jargon in its responses to the user.
As the conversation continues, the user's level of expertise may evolve.

\subsubsection{Sentiment ToM Helper:} After each user utterance, the \tool also sends the entire dialogue to the Sentiment ToM Helper. It monitors the sentiment flow of the discussion within the Human-GA conversation, categorizing it as ``Negative'', ``Neutral'', or ``Positive''. Based on such estimated sentiment, the Requirement Refiner Agent can adjust its discussion with the user (e.g., moving forward to the next question, rephrasing the topic, and/or changing the tone of the response).

\subsubsection{Extendability of Helpers:} The implementation of our ToM-based architecture within \tool provides several advantages regarding extensibility. Specifically, organizations may require domain-specific ToM helpers to assist the requirement refinement process by providing domain knowledge unique to each business.
\tool can support the seamless integration of such helpers as plugins or extensions. 
This plugin architecture allows organizations to flexibly choose one or more ToM helpers and accommodate future ones extending \tool capabilities. Additionally, there is a clear separation of concerns between the various ToM helpers, which enhances their maintainability. That also enforces the concept of specialized agents as discussed in Section~\ref{sec:concepts}.

\section{Evaluation Design and Results}
\label{sec:eval_results}

\begin{figure*}[h]
    \includegraphics[trim={0cm 0cm 0cm 0.2cm},clip,width=\textwidth]{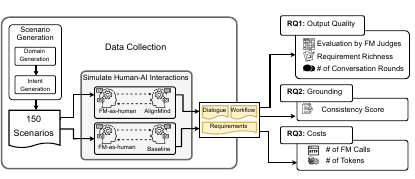}
    \caption{An overview of the evaluation of \tool. 
    }
    \label{fig:evaluation}
\end{figure*}

Figure~\ref{fig:evaluation} provides an overview of the evaluation for \tool.
This evaluation consisted of two main stages.
We first collected data to conduct the evaluation by simulating the human-AI conversations (Section~\ref{sec:eval_data_collection}).
Then, based on the collected data, we carried out the evaluation from three perspectives, with each perspective focusing on a particular research question. 

\subsection{Data Collection for Evaluation}
\label{sec:eval_data_collection}

Collecting data from the real world to evaluate the quality of a conversational agent in the domain of requirement refinement is challenging. Therefore, we opted to generate synthetic data using FMs to simulate human-AI interactions for the specific use case of requirements refinement, following similar approaches from prior studies~\cite{ge2024, abbasiantaeb2023let, argyrou2024automaticgenerationfashionimages, li2023camel, maronikolakis2024iwearpartygreek, zhou2024sotopia}. Thus, we create our dataset using the following steps:

\par\smallskip\noindent\underline{\textit{Scenario Generation.}} We adopt a multi-step approach to construct 150 scenarios across various domains with the goal of creating 150 dialogues between a pair of FMs. We first prompt an FM to generate ten diverse domains for which automated workflows can be developed. The complete system prompt of this agent can be found in the appendix~\ref{prompt:domain-gen}.
    Each domain is then combined with one of three personas (i.e., ``casual'', ``indecisive'', and ``rude''~\cite{maronikolakis2024iwearpartygreek}), and a random expertise level (i.e., ``novice'', ``intermediate'', or ``expert''), to obtain 30 configurations. 
    After that, we generate five intents for each configuration using a template-based technique, similar to prior work~\cite{maronikolakis2024iwearpartygreek}. This technique uses dynamic placeholders to diversify the generated dataset. These placeholders enable the generation of a diverse set of dialogues. Specifically, we consider four variables: (1) the expertise level of the user, (2) the persona of the user, (3) the domain,  and (3) one of two sentence fragments typically used to express an intent: `I would like to' or `I am looking for a way to.' Finally, we prompt an FM to complete the dynamically constructed sentence fragment with an intent. The template we used to generate the intents is given as follows: ``\texttt{As \{\{expertise\_level\}\} in \{\{domain\}\}, \{\{verb\}\}}''. For example, after following this process, one scenario generated through the system will be ``As a novice in Artificial Intelligence, I am looking for a way to receive bi-weekly notification of upcoming AI conferences and workshops, sent to my calendar through an automated API service.'' We obtain 150 different scenarios after following this process ($10 \times 3 \times 5 = 150$). 
    
\par\smallskip\noindent\underline{\textit{Simulate Human AI-Interactions.}} In this step, we use an FM-powered agent role-playing as a human to interact with \tool and the baseline based on each scenario generated in the previous step.
    The system prompt of the FM agent roleplaying as the human is specified in the Appendix~\ref{prompt:fm-as-human}.
    As the baseline, we use an FM-powered agent that is only instructed in the system prompt to refine user requirements. We included the full system prompt used in this agent in the Appendix~\ref{sec:baseline-prompt}.
    Based on each scenario, we invoke human-\tool and human-baseline conversations to generate a dialogue, a refined set of requirements, and a step-by-step workflow in natural language to achieve the requirements. At the end of this stage, we store the tuples consisting of the dialogue, requirements, and workflow of all 150 scenarios in a database for subsequent analyses.

In each subsequent section, we describe the motivation for studying the three research questions, outline our methodology to answer them using the generated data, and finally include the results.

\subsection*{RQ1 \rqi}
\label{sec:rq1}

We set out to investigate if \tool can improve the output of the requirement clarification and refinement process compared to directly prompting an FM, which we consider as the baseline, closely following the work of Wang et al.~\cite{wang2024xuatcopilotmultiagentcollaborativeautomated}. 
However, it is important to note that a single metric cannot fully capture the enhancements in the requirements refinement process or the quality of the artifacts produced by that process.
Therefore, we consider a multi-faceted approach based on the following three aspects:
\begin{itemize}
    \item \textbf{RQ1.1} Evaluation using a panel of FM-powered judges
    \item \textbf{RQ1.2} Evaluation of requirement richness
    \item \textbf{RQ1.3} Evaluation on the number of conversation rounds
\end{itemize}



\begin{figure*}[h]
    \includegraphics[trim={0cm 0cm 0cm 0cm},clip,width=\textwidth]{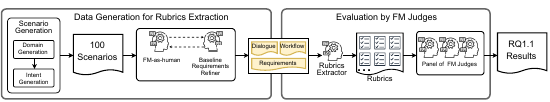}
    \caption{An overview of the output quality evaluation by a panel of FM judges (RQ1.1). 
    }
    \label{fig:eval_rq1_1}
\end{figure*}

\smallsection{RQ1.1  Evaluation using a panel of FM-powered judges}

\noindent\textbf{\textit{Motivation:}} As achieving consistent evaluation at scale with human evaluators is challenging, the "LLM-as-a-Judge" paradigm has emerged, in which LLMs are employed as evaluators for complex tasks~\cite{gu2024judgesurvey,li2024generationtojudgement}.
Even in the software engineering domain, Ahmed et al.~\cite{ahmed2024} show that replacing some human annotation effort with LLMs can produce inter-rater agreements equal to or close to human-rater agreement. Therefore, we chose to score the artifacts produced during the requirements refinement process (i.e., each tuple of dialogue, requirements, and workflow) by prompting a panel of three FM-powered judges. However, previous research has shown that using the same FM or closely related variants as evaluators can introduce bias due to preference leakage~\cite{li2025preferenceleakagecontaminationproblem} or self-preference~\cite{panickssery2024llmevaluatorsrecognizefavor}. 
They observed that an FM serving as both a predictor and an evaluator assigns disproportionately high ratings to its outputs compared to other FMs.
Therefore, to mitigate this bias, building upon recent work~\cite{verga2024} which showed the effectiveness of using a panel of FM evaluators, we chose three distinct FMs, drawn from two different model families: \llama from Meta and \gptmn, as well as \gpt from OpenAI, as our FM-powered judges.

Furthermore, we used single-point scoring~\cite{kocmi-federmann-2023-large} where the judge model is tasked with rating the quality of an output based on natural language instructions on how the grading should be performed (i.e. what properties constitute a good or bad output).
In our case, this requires providing the judging panel with clear evaluation criteria, or rubrics, that are relevant to the process of refining requirements through conversation. To the best of our knowledge, no such guidelines currently exist.
However, prior work~\cite{Biyani2024, lin2024interpretable} has used conversations conducted with an FM-powered agent to generate a set of rubrics to estimate domain-specific user satisfaction in other domains such as software debugging. 
Inspired by this work, we opted to generate synthetic data using FMs to simulate human-AI interactions for the use case of requirements refinement and derive rubrics from this data.

\noindent\textbf{\textit{Approach:}} Figure~\ref{fig:eval_rq1_1} illustrates the approach we followed to evaluate \tool using the panel of FM-powered judges.
The process consists of two stages: (1) Data generation for rubrics extraction and (2) evaluation by the FM-powered judge panel, which we describe next. 

\par\smallskip\noindent\underline{\textit{Data generation for rubrics extraction.}} Similar to the steps in Section~\ref{sec:eval_data_collection}, we first generate a set of 100 scenarios with the help of an FM. For this purpose, we begin by prompting an FM to generate 20 diverse domains for which automated workflows can be developed. After that, we generate five diverse intents for each domain using the same template-based technique.

Based on these 100 domain and intent combinations (i.e., 100 scenarios) we use a pair of FM-powered agents to generate a dialogue. The first FM-powered agent, roleplaying as a human, is provided with each one of the 100 scenarios and prompted to initiate the conversation. The second FM agent serves as a baseline requirement refiner agent, helping the former agent clarify its intent. The refiner agent terminates the dialogue when it has collected sufficient information to achieve the user's intent. The specific system prompts used in the FM agent roleplaying as the human and requirements refiner agent are available in Appendix~\ref{prompt:fm-as-human} and Appendix~\ref{prompt:fm-as-refiner} respectively.

Once the dialogue is terminated, an FM is prompted to extract the refined requirements from the entire dialogue into a requirements document. Then, another FM call is used to create a step-by-step workflow in natural language based on the requirements document. Each combination of the dialogue, requirements, and workflow for all 100 scenarios is passed on to the next step for rubric extraction and evaluation by FM judges.

\par\smallskip\noindent\underline{\textit{Evaluation by FM-powered judges.}} Inspired by prior work~\cite{Biyani2024, lin2024interpretable}, first, we use FMs to derive rubrics to assess the quality of the generated artifacts during the requirements refinement. We employ a two-step process to generate rubrics.  First, we provide a combination of dialogue, requirements, and workflow to an FM-powered agent to extract three reasons why these artifacts might be considered good. The full system prompt used can be found in the Appendix~\ref{prompt:rub-reasons-gen}. This process yields a total of 300 reasons, with three reasons derived from each of the 100 data points. Next, we prompt another FM-powered agent to generate rubrics based on the extracted reasons. The full system prompt of this agent is available in the Appendix~\ref{prompt:rub-gen}. After manual inspection to remove duplicates, the list of five rubrics, which was produced by this process, is presented below: 
\begin{Summary}{Rubrics}{}
\begin{itemize}
    \item[$\blacksquare$] The assistant is able to accurately identify the user's intent.
    \item[$\blacksquare$] The requirements capture all of the user’s intent with respect to their requirements, preferences, and perceptions.
    \item[$\blacksquare$] The requirements are relevant to achieve the user’s intent.
    \item[$\blacksquare$] The workflow includes detailed, actionable, and ordered steps.
    \item[$\blacksquare$] The workflow is realizable and error-free.
\end{itemize}

\end{Summary}

Next, based on these rubrics, we evaluate the artifacts (dialogue, requirements, and workflow) generated by \tool and the baseline solution described in Section~\ref{sec:eval_data_collection}.
For this purpose, We employ three FMs, drawn from two different model families: \llama from Meta and \gptmn, as well as \gpt from OpenAI, as our panel of FM-powered judges for the evaluation based on the five rubrics.
    The full system prompt used in each of these evaluator agents to rate each triplet is available in the Appendix~\ref{prompt:rub-judge}.
    To improve the robustness of the result~\cite{castillo2024}, we prompt the evaluator to provide reasoning before assigning a score using a 5-point Likert scale. This 5-point score, which ranges from \textit{Strongly Disagree} to \textit{Strongly Agree}, is converted into a normalized score within the [0, 10] range, where 0 indicates Strongly Disagree, 2.5 indicates Disagree, 5 indicates Neutral, 7.5 indicates Agree, and 10 indicates Strongly Agree.  To ensure the consistency of our evaluation results, we prompt each evaluator agent three times and compute the mean for each rubric score. The overall score for a given triplet is calculated as the mean of all rubric scores as follows:
    \[
\text{Overall Score}(D, R, W) = \frac{1}{N} \sum_{i=1}^{n} C_{DRW}(i)
\]

    \begin{itemize}
        \item \(D\) : Dialogue.
        \item \(R\) : Generated Requirements.
        \item \(W\) : Generated Workflow.
        \item \(C_{DRWn}\)  : Score of the \(n\)-th rubric.
        \item \(N\) : Total number of rubrics.
    \end{itemize}

 Then, we use the Wilcoxon signed-rank test to determine if there is a statistically significant difference between the distributions of scores between the baseline and \tool. 
Moreover, we use Cliff’s delta to measure the effect size, which
is negligible when \textit{delta} $<$ 0.147 , small when 0.147 $\leq$ \textit{delta} $<$
0.33 , medium when 0.33 $\leq$ \textit{delta} $<$ 0.474 , and large otherwise.
Furthermore, we use the median value for aggregating scores across the panel of three judges. (e.g., For a given scenario out of 150, if the three judges have given a higher median overall score to the \tool, compared to the baseline, \tool is chosen as the preferred approach for that scenario.)

To validate whether FM-powered judges are aligned with human judgment, we compare the preference of the FM judge panel with the human-provided preference for a subset of 20 scenarios. 
We assess inter-rater reliability using Cohen’s $\kappa$ coefficient~\cite{Emam1999}, which measures the degree of
agreement between the FM-chosen labels and human labels.
We obtain a value of 0.685 for $\kappa$, indicating substantial agreement, and therefore continue with the evaluation.

\begin{figure}[h]
    \includegraphics[width=0.5\textwidth]{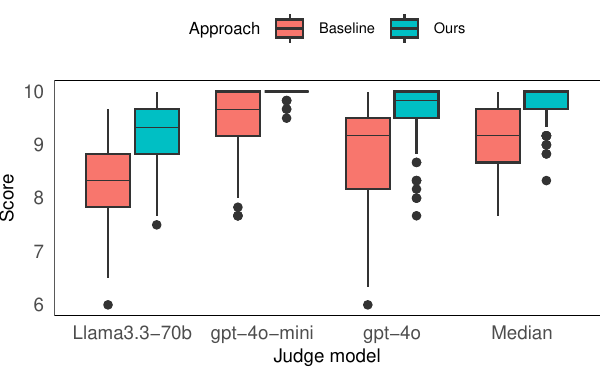}
    \caption{The overall score of \tool vs baseline, as judged by three different FMs. The rightmost plot (Median) shows the distribution of scores where, for each scenario, the median score from the three judge models is taken. 
    }
    \label{fig:overall-scores}
\end{figure}

\begin{figure}[h]
    \includegraphics[width=0.5\textwidth]{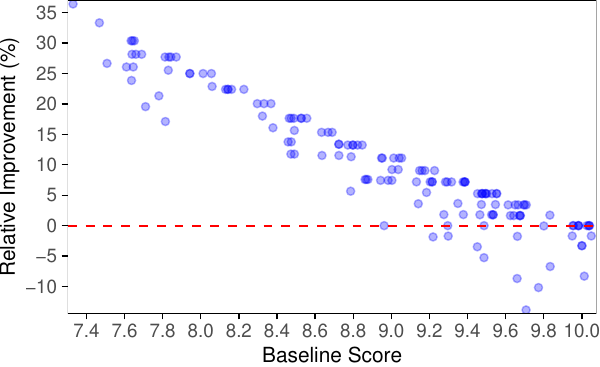}
    \caption{Scatterplot showing the relative improvement of \tool, compared to the baseline, across 150 different scenarios. Data points that lie above the red line indicate instances where \tool improves the overall score compared to the baseline.}
    \label{fig:solution-improvement}
\end{figure}

\noindent\textbf{\textit{Results:}} Figure~\ref{fig:overall-scores}
illustrates the performance comparison between the baseline and \tool based on the overall evaluation score.


\textbf{\tool outperforms the baseline in the evaluations by all three FM-powered judges, with the overall score differences being statistically significant.} Specifically, incorporating \tool's capabilities in requirements clarification and refinement tasks results in a higher overall performance score, as illustrated in Figure~\ref{fig:overall-scores}. Notably, different judge models exhibit consistent performance trends across both \tool and the baseline configurations. The Wilcoxon signed-rank test reveals statistically significant results favoring \tool in evaluations by \llama, \gptmn, and \gpt, with p-values of \num{1.95e-20}, \num{2.42e-18}, and \num{1.45e-14}, respectively. 
The median of the distribution of aggregate scores across all three judge models for the baseline and \tool are 9.08 and 10, respectively.

Figure~\ref{fig:solution-improvement} shows the relative improvement in aggregate scores when \tool is used, compared to the baseline, across different scenarios. The relative improvement ranges between -13.85\% and 36.42\% with the median at 7.44\%. 122 out of 150 (81.33\%) scenarios show an improvement in the overall score with \tool compared to the baseline.

\textbf{\tool is particularly useful in situations where the baseline approach is facing challenges.} According to Figure~\ref{fig:solution-improvement}, the scenarios that received lower overall scores with the baseline approach exhibit the greatest relative improvement when \tool is used. This clearly highlights the effectiveness of our method in tackling complex scenarios.

\begin{figure}[h]
    \includegraphics[width=\textwidth]{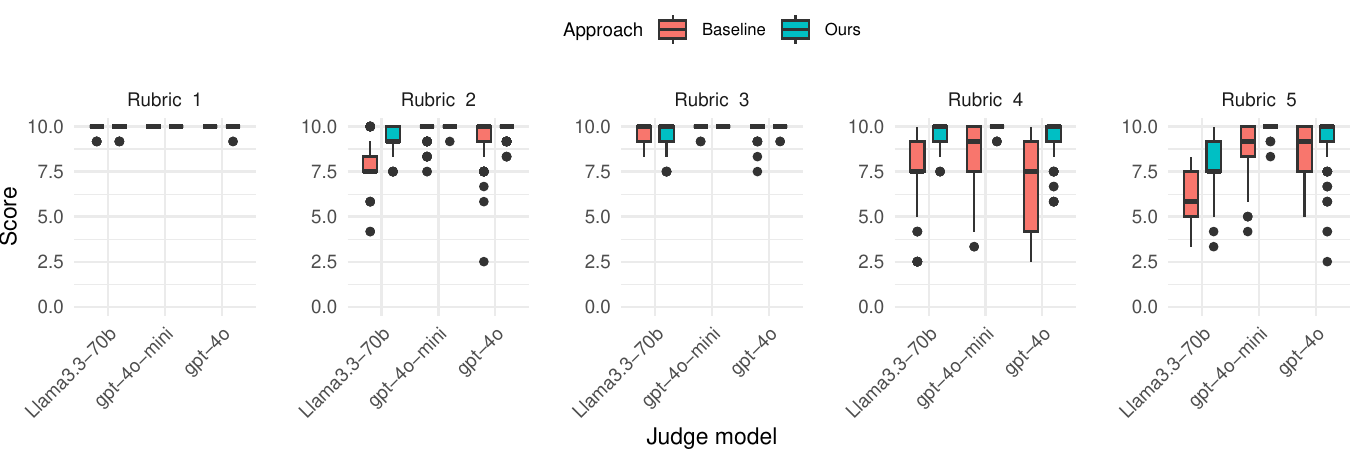}
    \caption{The score for each rubric of \tool and the baseline as judged by three different FMs.}
    \label{fig:rub-score}
\end{figure}

\textbf{When the individual rubrics are considered, \tool is unanimously chosen as better in three out of five rubrics.} We observe that all the considered judge models yield better performance in most of the rubrics, as shown in Figure~\ref{fig:rub-score}. In particular, when using \llama and \gptmn as the judge models, \tool achieves better performance in rubrics 2, 4, and 5, with statistically significant differences (as illustrated in Table~\ref{tab:rub-stats}). Furthermore, \gpt favors \tool in four out of five rubrics (2 to 5). Interestingly, we notice that \tool and the baseline exhibit similar performance for rubric 1 across all judge models. Only in one case (when even evaluating rubric 3 using \llama judge model), the baseline has outperformed \tool. Hence, our results suggest that the obtained performance measures are reliable across different judge models. Moreover, \tool, supported by its advanced clarification features, demonstrates promising results in the requirements refinement task.

\begin{table}[ht]
\centering
\caption{Wilcoxon Test Results for each rubric when three different FMs are used as judges. \textit{Win.} stands for Winner. \textit{Sign.} stands for Significant.}
\label{tab:rub-stats}\begin{adjustbox}{width=\textwidth}
\begin{tabular}{lcccccccccccc}
\toprule
\textbf{Rubric} & \multicolumn{4}{c}{\textbf{Llama3.3-70b}} & \multicolumn{4}{c}{\textbf{gpt-4o-mini}} & \multicolumn{4}{c}{\textbf{gpt-4o}} \\
\cmidrule(l){2-13} & p-value & Win. & Sign. & Cliff's delta & p-value & Win. & Sign. & Cliff's delta & p-value & Win. & Sign. & Cliff's delta \\
\midrule
Rubric  1 & \num{ 1.00e+00 } & N/A & No  &  -0.00  & \num{1.00e+00} & N/A & No  &  0.00  & \num{ 3.46e-01 } & N/A & No  &  -0.01  \\
Rubric  2 & \num{ 8.16e-19 } & AlignMind & Yes  &  0.68  & \num{5.28e-06} & AlignMind & Yes  &  0.17  & \num{ 5.92e-08 } & AlignMind & Yes  &  0.28  \\
Rubric  3 & \num{ 9.51e-04 } & Baseline & Yes  &  -0.17  & \num{1.49e-01} & N/A & No  &  0.02  & \num{ 2.47e-02 } & AlignMind & Yes  &  0.05  \\
Rubric  4 & \num{ 1.59e-20 } & AlignMind & Yes  &  0.69  & \num{5.49e-17} & AlignMind & Yes  &  0.60  & \num{ 1.10e-17 } & AlignMind & Yes  &  0.65  \\
Rubric  5 & \num{ 7.06e-18 } & AlignMind & Yes  &  0.67  & \num{2.69e-16} & AlignMind & Yes  &  0.57  & \num{ 3.41e-05 } & AlignMind & Yes  &  0.28  \\
\midrule
\begin{tabular}{@{}c@{}}Overall \\ Score\end{tabular}
& \num{ 3.94e-47 } &  AlignMind & Yes & 0.29 & \num{ 1.51e-36 } &  AlignMind & Yes & 0.27 & \num{ 1.02e-27 } &  AlignMind & Yes & 0.22 \\
\bottomrule
\end{tabular}
\end{adjustbox}
\end{table}


From the practical point of view, \tool can typically capture all users' intent concerning their requirements, preferences, and perceptions (rubric 2). Furthermore, \tool can help achieve the user's intent (rubric 3). Additionally, \tool generates a detailed, actionable, and ordered-step workflow (rubric 4), while such a workflow is realizable and error-free (rubric 5).

\begin{figure}[h]
    \includegraphics[width=0.5\textwidth]{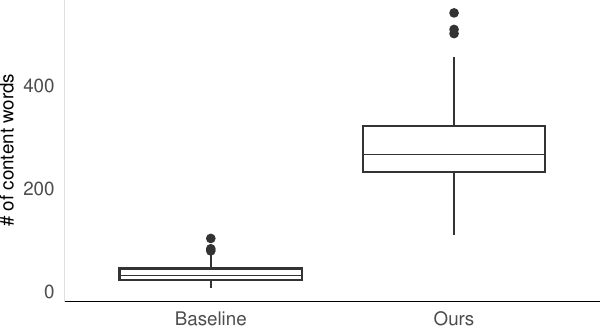}
    \caption{The requirements richness between \tool and baseline configurations. The median of the baseline approach is 33 and \tool is 266.5.}
    \label{fig:req-richness}
\end{figure}


\smallsection{RQ1.2  Evaluation of requirement richness}

\noindent\textbf{\textit{
Motivation:}}
In this RQ we set out to asses the quality of the generated requirements through an objective measurement.
In previous work~\cite{Zhao2021, Sonbol2022}, lexical features of natural language have been used to extract requirements from documents. Therefore, to assess the richness of the generated requirements, drawing inspiration from this previous work, we compare the lexical diversity~\cite{kyle2019measuring}, specifically the number of unique and non-stop words in the final refined requirements output by both \tool and the baseline. 

\noindent\textbf{\textit{
Approach:}}
For this purpose, we begin by removing stop words from the refined requirements.
Next, we transform the requirements into a matrix of tokens, with each token representing a word using the \texttt{CountVectorizer} from the \texttt{scikit-learn} Python library. The result is a binary matrix where each element indicates the presence of a specific word. Using this matrix, we then count the number of unique content words.

\noindent\textbf{\textit{Results:}} Figure~\ref{fig:req-richness} illustrates the richness of requirements generated by \tool and the baseline, as measured by the number of content words.

\textbf{Requirement richness achieved by \tool is eight times higher than that of the baseline.} 
The median number of content words in the final set of requirements when the \tool is used for requirements refinements is 266.5, while it is 33  for the baseline. 
With its advanced capabilities, \tool generates richer content in the final requirements compared to the baseline approach. A Wilcoxon signed-rank test confirms that this difference is statistically significant, with a p-value of \num{2.31e-26}.
The cliff's delta is 1, suggesting that the requirements richness output by \tool in all instances is higher than the outputs in the baseline approach, with no overlapping values between the two groups.
These results suggest that employing a multi-agent architecture supported by various refinement features, such as those in \tool, can be highly effective in enhancing requirement richness.


\smallsection{RQ1.3  Evaluation on the number of conversation rounds}

\noindent\textbf{\textit{Motivation:}} Previous work~\cite{Bajpai2024,wu2024} has identified that FMs tend to prematurely terminate conversations by jumping straight to providing solutions, even when the information provided by the user is insufficient to reach a solution.
Therefore, we set out to investigate how effectively \tool can mitigate this weakness by extending conversations with the user when necessary, with the support of our proposed improvements.

\noindent\textbf{\textit{Approach:}} 
To address this RQ, we compare the number of conversation rounds maintained by \tool and the baseline solution when interacting with the FM agent role-playing as a human while refining requirements for the 150 scenarios.

\begin{figure*}[h]
    \centering
    \includegraphics[width=0.5\textwidth]{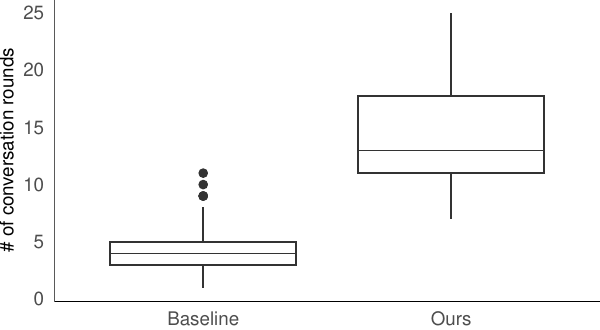}
    \caption{The number of conversation rounds used for requirements refinement using \tool vs. baseline.}
    \label{fig:num-rounds}
\end{figure*}

\noindent\textbf{\textit{Results:}} As depicted in Figure~\ref{fig:num-rounds}, \tool engages in a median of 13 conversation rounds with the user, whereas the baseline approach results in a median of only four rounds. 

\textbf{\tool enables longer multi-round conversations compared to the baseline.}
Even when the system prompt does not explicitly instruct to ask clarifying questions, the baseline approach still has four conversation rounds.
However, this is significantly lower than the 13 conversation rounds of \tool.
The Wilcoxon signed-rank test confirms a statistically significant difference with p-value = \num{2.14e-26}. We also obtain a Cliff's delta of 0.986, suggesting a large effect size. This means that \tool overwhelmingly outperforms the baseline approach in terms of guiding users through several refinement steps to generate a refined set of requirements and a natural language workflow.
A higher number of conversation rounds allows for deeper clarification of user intent through iterative dialogue rather than relying on a single direct prompt to the FM for the same task~\cite{wu2024}. This difference highlights \tool’s ability to guide users through a more comprehensive and iterative refinement process, ensuring grounded and contextually relevant questions across various requirement aspects. 

\begin{Summary}{Summary of RQ1}{}
\keyfinding{\tool demonstrates superior performance in terms of overall score, with a statistically significant difference compared to the baseline. Furthermore, we observe that \tool outperforms the baseline in three to five rubrics across the judge models, as confirmed by the Wilcoxon signed rank test. Similarly, \tool significantly improves requirement richness compared to the baseline with a reasonable number of conversation rounds.}

\implications{Our solution, \tool, can assist software developers and non-technical professionals across various organizations in requirements refinement, enabling the creation of rich and well-structured requirements and workflows. Furthermore, our results suggest that the integration of clarification capabilities, such as those in \tool, can be valuable for requirements refinement tasks.}
\end{Summary}
\smallsection{RQ2 \rqii}

\noindent\textbf{\textit{Motivation:}} While RQ2 touches on how much information the summarized requirements contain, content alone may not serve as a reliable metric, as it could include hallucinated information. Therefore, in this research question, we investigate the hallucination aspect of the summarized requirements (the extent to which the summarized requirements are consistent with and specific to the dialogue).

\noindent\textbf{\textit{Approach:}} 
Kryscinski et al.~\cite{Kryscinski2019} proposed four dimensions for evaluating abstractive summaries: relevance, coherence, consistency, and fluency.
Out of these dimensions, consistency, which checks if the summary aligns with the facts in the source document, is the most relevant for hallucination detection.
The intuition is that the summary and source document should be factually consistent if no new information is present in the summary that’s not in the source. 
Therefore, we formulate our problem of checking the final requirements for hallucination as a problem of consistency checking in abstractive summaries. 
In our context,  the human-AI conversation is considered the source document, while the final set of requirements is considered the summary.

We follow OpenAI's guidelines and use FMs-as-judges for determining consistency in the summaries.~\footnote{\url{https://cookbook.openai.com/examples/evaluation/how\_to\_eval\_abstractive\_summarization\#evaluating-using-gpt-4}} 
We use the same three models as judges that we employed in the previous RQ: Llama3.3-70b, gpt-4o-mini, and gpt-4o. The system prompt includes guidelines for evaluating the consistency between a specified source document and its summary. Additionally, we ask the judging model to assign a consistency score ranging from 0 to 5.

For each of the 150 scenarios, we provide the dialogues and the refined requirements generated by the baseline and \tool, independent of each other, and prompt each of the judge FMs to provide a consistency score. Then, we use the Wilcoxon signed-rank test to determine if there is a statistically significant difference between the distributions of consistency scores between the baseline and \tool. 
Moreover, we use Cliff’s delta to measure the effect size.

\noindent\textbf{\textit{Results:}} \textbf{All three judge models gave full score (5 out of 5)  for consistency, for output generated by both baseline and \tool in the majority of the cases.} The median consistency score was 5 for both baseline and \tool by all judge models. This demonstrates that requirements generated by both baseline and \tool are consistent with the AI-human conversations, which were used for requirements refinement (i.e, no additional requirements not discussed in the user conversations were included in the final set of requirements due to hallucination).

\textbf{No statistical difference between the consistency scores of baseline and \tool, based on assessment by two out of three FM-based judges.} 
Based on the Wilcoxon signed-rank test results for Llama3.3-70b and gpt-4o-mini judge models, there was no evidence to claim a statistical difference in consistency in output generated by baseline and \tool. When GPT-4o is used as the judge model, based on the Wilcoxon signed-rank test, \tool consistency scores are significantly higher than the baseline ( p-value \num{ 1.71e-06 }). Moreover, we obtain a Cliff’s delta of 0.19, which is considered small but non-negligible.

\begin{Summary}{Summary of RQ2}{}
\keyfinding{Both the \tool and the baseline produce hallucination-free requirements from the requirements clarification and refinement process, consistent with the user conversations.}

\implications{\tool proves to be an effective tool towards creating rich, high-quality requirements specifications that are grounded in dialogue. Moreover, \tool has the potential to be a useful resource for various stakeholders, not just in refining requirements but also in areas such as requirements elicitation.}
\end{Summary}
\smallsection{RQ3 \rqiii}


\noindent\textbf{\textit{Motivation:}} This research question seeks to estimate the costs associated with using an FM-based conversational agent system enhanced with ToM capabilities for requirement refinement tasks. Specifically, we aim to evaluate whether \tool can assist organizations in clarifying requirements efficiently while maintaining reasonable costs. The findings of this study could provide valuable insights for various industries, helping them better understand the cost implications of using a multi-agent solution vs. a single agent in requirements refinement tasks.

\noindent\textbf{\textit{Approach:}} We quantify the expenses associated with using our \tool vs. the baseline for requirement refinement tasks. To address this research question, we consider three different cost-related metrics. Below, we provide a brief introduction to each of these metrics:

\begin{itemize}
    \item \textbf{Number of FM calls}: This metric represents the number of calls that \tool performs to foundation models during a session used for requirements refinement. Since FM calls are usually made over HTTP to externally hosted foundation models, this metric gives an estimate of network resource usage and latency while \tool is in operation at an organization.
    \item \textbf{Number of tokens}: All major FM hosting providers currently charge users based on token counts. This includes both input (aka \textit{prompt\_tokens}) and output (aka \textit{completion\_tokens}). Therefore, by calculating token usage, we can estimate the financial costs of using \tool. OpenAI provides a lightweight method to measure the cost of tokens consumed in each FM call. Typically, when invoking any OpenAI model, three key pieces of information are returned from the endpoint: \textit{prompt\_tokens}, \textit{completion\_tokens}, and \textit{total\_tokens}. Since we invoke OpenAI models multiple times for each intent-based instance, we sum each of these token metrics across all FM calls made for each use case.
\end{itemize}

\begin{figure}[h]
    \centering
    \includegraphics[width=.5\textwidth]{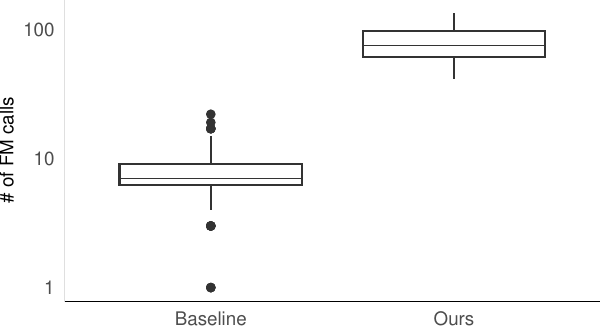}
    \caption{The number of FM calls required in the two configurations for requirements refinement.}
    \label{fig:http-requests}
\end{figure}

\noindent\textbf{\textit{Results:}} \textbf{The \tool version uses a higher number of FM calls compared to the baseline.} 
Our findings indicate that performing requirements refinement tasks with \tool requires a median of 74.5 calls to FMs, while the baseline necessitates only 
seven requests, as shown in Figure~\ref{fig:http-requests}. The Wilcoxon signed-rank test shows a statistically significant difference with a p-value of \num{2.30e-26}.
Every observation generated by \tool has a higher number of API calls than every observation in the baseline.
We attribute these observations to the multi-agent architecture of the \tool, which employs several FM-based agents to facilitate the requirements refinement process. Consequently, a higher number of FM calls is expected. 
Each interaction begins with a pair of FM agents that collaboratively decompose the user's intent into a set of topics and subsequently prompt another FM to generate relevant questions. 
Additionally, \tool leverages a range of ToM-based agent helpers to improve the requirement refinement. Such helpers are invoked after each user query. 

\begin{figure*}[h]
    \centering
    \begin{subfigure}[b]{0.32\textwidth}
        \includegraphics[width=\textwidth]{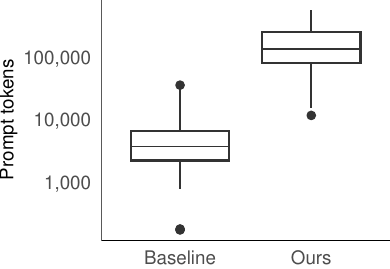}
        \caption{Prompt tokens.}
        \label{fig:prompt-tokens}
    \end{subfigure}
    \hfill
    \begin{subfigure}[b]{0.32\textwidth}
        \includegraphics[width=\textwidth]{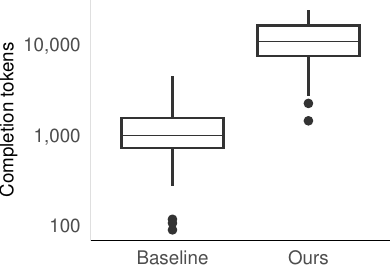}
        \caption{Completion tokens.}
        \label{fig:completion_tokens}
    \end{subfigure}
    \hfill
    \begin{subfigure}[b]{0.32\textwidth}
        \includegraphics[width=\textwidth]{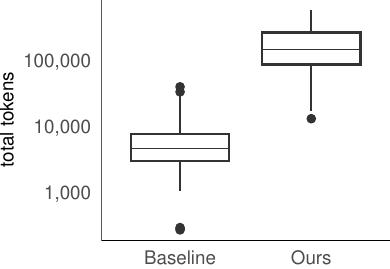}
        \caption{Total tokens.}
        \label{fig:total-tokens}
    \end{subfigure}
    \caption{The cost in terms of (a)~\textbf{prompt}, (b)~\textbf{completion}, and (c)~\textbf{total tokens} needed for requirements refinement in the baseline configuration vs \tool.}
    \label{fig:token-cost}
\end{figure*}

\textbf{There is a notable monetary cost associated with using \tool compared to the baseline, as measured by metrics related to prompt (input), completions (output), and total tokens.} Our findings indicate that \tool incurs a higher token cost compared to the baseline across all metrics, as illustrated in Figure~\ref{fig:token-cost}. Specifically, the median token usage is as follows: 129,181.5 for prompt tokens, 10,139 for completion tokens, and 139,784 for total tokens with \tool, contrasted with 3,758.5 for prompt tokens, 981.5 for completion tokens, and 4,735 total for tokens for the baseline. A Wilcoxon signed-rank test shows a statistically significant difference with p-values \num{7.59e-26}, \num{1.11e-25}, and \num{7.75e-26} for prompt, completion, and total tokens, respectively. All Cliffs' Delta values are 0.87, which is considered a large effect size. This suggests almost no overlap between \tool and the baseline in token usage. Accordingly, this increase in token usage aligns with our earlier observations, as it is reasonable to expect a higher cost due to the multi-agent architecture of our solution. Each FM call necessitates sending the entire conversation history, including the prompt, to maintain a coherent dialogue and ensure accurate responses. Moreover, given the multi-agent structure of our solution, one can anticipate an increase in tokens, particularly in the prompt tokens providing instructions for each agent. 
It's important to recognize that not all of the 10,139 output tokens will be visible to the end user. Most of these tokens are used by FM-powered agents for internal decision-making processes. Therefore, while overwhelming users with the sheer number of tokens is not a concern, we should still be mindful of the financial implications since costs for FM access are currently calculated based on the number of tokens consumed and generated.
Another point is that prompt tokens are generally priced lower than completion tokens. Our method tends to use more input tokens than output tokens, so the cost may not be as dire as it first appears.\footnote{\url{https://openai.com/api/pricing/}} 

\textbf{Despite the cost difference between \tool and the baseline, \tool remains both cost-effective and efficient for requirements refinement tasks compared to manual requirement refinement. }
For instance, the operational costs are approximately a median of \$0.94 using Gemini-1.5-flash (Google), \$0.97 using Claude-3-haiku (Anthropic), and \$31.40 using \gpt (OpenAI) for requirements refinement tasks with \tool, while it costs as low as \$0 and as high as \$0.13 for the same FM providers using the baseline approach.\footnote{Based on the prices on December 5th, 2024 
 as listed on \url{https://gptforwork.com/tools/openai-chatgpt-api-pricing-calculator}} With inference costs further decreasing~\cite{eladtweet} and techniques like prompt caching~\cite{humanloopPromptCaching} and prompt compression~\cite{Jiang2023} being supported by major FM service providers, we postulate that FM-based solutions will become an attractive choice for requirement refinement and intent alignment in the coming years.

\begin{Summary}{Summary of RQ3}{}
\textbf{Findings}: Our findings underscore notable computational and operational overhead when leveraging \tool for requirement refinement. In particular, our solution requires a median of 74.5 FM calls, while it costs as low as \$0.97 and as high as \$31.40 according to various FM providers.

\textbf{Implications}: Our analysis offers key insights for various corporate stakeholders regarding the computational and operational costs associated with leveraging \tool multi-agent solution to refine requirements. With the inference costs getting reduced further, we anticipate the emergence of numerous innovative solutions based on the multi-agent architecture and various ToM capabilities, which will significantly revolutionize the requirements engineering field, mainly requirement refinement.
\end{Summary}

\section{Discussion and Implications}
\label{sec:discussion}

From the experience of building \tool and trialling it out in the field, we have learned some lessons and have seen a glimpse of what the future holds in intent alignment and requirements refinement with the assistance of FMs. 

\smallsection{Assisting with the Performance-Cost Tradeoff.} Compound AI systems and agentic AI systems such as \tool are complex and known to be costly~\cite{Hassan2024,Kapoor2024}. In contrast to a single model call, agent-based systems call the models multiple times before arriving at the final solution. Based on current pricing plans offered by model hosting providers, costs are incurred each time the agent is run and depend on the number of input and output tokens. Based on the use case, developers may decide to compromise accuracy to reduce costs. Tools must be provided for the FM-based application developers to visualize these factors and manage the tradeoff between cost and accuracy. Furthermore, they may benefit from techniques~\cite{dspy} that can optimize costs by reducing the prompt length, output format, and number of agent hops while maintaining functionality and overall performance.

\smallsection{The multi-modal future.} Requirements engineering is a quintessentially multimodal process involving a variety of modes and media to gather, articulate, and refine the needs and expectations of stakeholders in a software development project~\cite{yang2024swebenchmultimodalaisystems}. Stakeholders engage in discussions to express needs and expectations. These verbal conversations can occur in formal meetings, workshops, or informal dialogues, with different levels of precision. Furthermore, diagrams, charts, and sketches crafted on whiteboards or other digital tools occasionally provide visual representations of system components, workflows, or user interactions. Written documents, including emails, requirements specifications, and notes, serve as permanent records and reference points, capturing details and justifications for each requirement over long periods. Therefore, AI-based requirements refinement should get support beyond large language models and integrate with multi-modal foundation models. In future work, we plan to address this challenge.

\smallsection{New UX Paradigms.} The chat UI may not be the optimal way to interact with foundation models to achieve complex tasks in domains such as software development. On the other hand, organizations are experimenting with new ways for users to interact with foundation models. v0 by Vercel\footnote{https://vercel.com/blog/announcing-v0-generative-ui}, Artifacts by Anthropic\footnote{https://www.anthropic.com/news/claude-3-5-sonnet}, and Townie\footnote{https://blog.val.town/blog/codegen/} are one such category of interfaces where interactive UI components are rendered dynamically from FM output. Users can refine these FM-generated interfaces as needed and copy them into their own projects. It is uncertain at this point in time, in a future where AI systems are expected to be intelligent collaborators, which of these experiments will gain traction~\cite{hassan2024ainative}. We are confident that the features provided by \tool to preview the captured requirements and generated workflow in real-time, as well as to modify them using natural language, also offer a glimpse of the future of intent-first IDEs.

\smallsection{Beyond Software Domain.} The applications of a goal alignment agent that can help with intent alignment and requirements refinement are not limited to Software Engineering. It can benefit both technical and non-technical professionals by bridging domain-specific knowledge and communication gaps. For software developers, the agent assists by systematically capturing detailed software requirements, ensuring that the developed system aligns with stakeholder objectives and user needs, reducing the likelihood of costly rework and project delays. In the automotive sector, engineers can utilize the agent to clarify complex design specifications and safety standards, facilitating effective collaboration between multidisciplinary teams and ensuring compliance with industry regulations. For healthcare practitioners, the agent aids in translating clinical requirements into actionable software features, enhancing the development of medical informatics tools that improve patient care and data management. Analysts can leverage the agent to articulate complex financial models and compliance requirements in the financial sector, ensuring that the resultant software solutions accurately reflect regulatory standards and market dynamics. The conversational agent enhances cross-disciplinary communication and fosters a more integrated approach to problem-solving across various industries by providing a natural language interface for capturing and aligning diverse requirements.

\smallsection{Plug-and-play Architecture.} As the need for requirements refinement and intent alignment spans across industries, a Foundation-model-based goal alignment solution can serve a broad spectrum of industries, offering the ability to create innovative, intent-driven solutions. The flexibility provided by \tool plug-and-play architecture of \tool allows organizations to seamlessly develop custom modules tailored to their unique requirements and constraints. By leveraging Theory-of-Mind-based adapters, the solution ensures a deeper alignment with the perspectives of various stakeholders. However, recognizing and responding to user perceptions can be a complex task in practice~\cite{wang2021towards}. In particular, one might need to put some effort into creating a set of ToM-based helpers for a more aligned user's output.

\smallsection{Importance of Domain-Specific Rubrics.} Rubrics serve as a valuable foundation for evaluating the performance of foundation models (FMs), as seen in our work and in program debugging tasks~\cite{Biyani2024}. However, designing effective rubrics requires careful consideration to ensure consistency and domain specificity. For instance, Biyani et al.~\cite{Biyani2024} found that domain sensitization plays a crucial role in significantly improving final evaluation scores. Their rubrics, derived from a debugging-code dataset, highlight the importance of tailoring evaluation criteria to the nuances of a specific field. Additionally, industries should therefore prioritize the development of accurate, domain-specific rubrics to ensure comprehensive and precise assessment. Well-crafted rubrics enhance the reliability of evaluation outcomes, leading to more meaningful insights. This is especially critical in domains where human cognitive factors, such as intent and communication styles, play a key role.

\smallsection{Comprehensive Evaluation Using Diverse FMs.}
Many existing studies on leveraging foundation models (FMs) for software engineering tasks rely on the same FM as both predictor and evaluator. However, this approach can introduce critical biases, leading to phenomena such as preference leakage~\cite{li2025preferenceleakagecontaminationproblem} or self-preference~\cite{panickssery2024llmevaluatorsrecognizefavor}, where an FM tends to favor its own predictions over those of other models. To mitigate this issue, we employ three variants of FMs from two widely used families: \llama from Meta; \gptmn and \gpt from OpenAI. Our evaluation reveals slight variations in assessments among the three models. However, \gpt judge model consistently favors our proposed approach, \tool, over the baseline across all rubrics and summarization-based metrics. Consequently, our findings highlight the importance of using diverse FMs for evaluation. Therefore, researchers are encouraged to adopt this approach to enhance the consistency and comprehensiveness of FM performance evaluations.

\section{Threats to Validity}
\label{sec:threats}

\smallsection{Internal Validity} refers to the extent to which the observed effect is indeed due to the independent variable and not other factors. In our case, the inherently stochastic nature of LLMs can introduce variability in the responses generated at each intermediate step where an LLM call is made in the requirements refinement system. To mitigate the effects of this behaviour for our evaluation, we have queried the evaluator model three times for each instance and computed the mean score.

\smallsection{External Validity} refers to the extent to which the findings of the study can be generalized to other populations or settings. In this study, which focused on requirements refinement, we specifically used the Meta LLaMA 3.3 and OpenAI GPT-4o class of models for our implementation and evaluation. It's important to note that results could vary if models from other providers, like Google Gemini, Anthropic Claude, or Mistral, were employed. Nevertheless, recent benchmark findings indicate that the performance of models from all leading labs has begun to converge. This trend can likely be attributed to the similarities in model architectures and significant overlaps in pre-training data~\cite{benaich2024state}. Therefore, we hypothesize that the influence of varying models on the performance of \tool will be negligible, producing consistent results.

\smallsection{Construct Validity} concerns the extent to which the model measures the intended construct or concept. In our study, we adopt a set of rubrics to evaluate the quality of the conversations, requirements, and natural language workflows generated by FM-powered agents, similar to prior work~\cite{Biyani2024}. To avoid potential biases from solely using FM-generated rubrics, we carefully crafted rubrics to account for the primary purposes, requirements refinement and natural language workflow generation. Each rubric was then manually reviewed by the first two authors for relevance. The authors' analysis of each rubric and discussion resulted in the final set of five FM-generated rubrics mitigating bias.

\section{Conclusion}
\label{sec:conclusion}

A foundation-model-powered multi-agent requirement refinement system with Theory-of-mind capabilities can accurately capture stakeholder requirements and intentions. The refined requirements and the natural language workflows that operationalize and help validate them are of high quality in terms of FM-evaluated metrics and lexical diversity. 
Furthermore, the proposed solution is able to continue conversations with the user to clarify requirements in the presence of inconsistencies, ambiguities, and incompleteness, without prematurely jumping to solutions.
The requirements produced through this process are grounded in the conversations and not affected by hallucinations.
However, the proposed approach uses significantly more tokens and API calls for requirements refinement compared to the baseline.
Informed by these observations, we hypothesize that these results will lead to future innovation in building intent-first development environments.

\bibliographystyle{ACM-Reference-Format}
\bibliography{10-references}

\appendix
\section{Appendix}
In this section, we present different prompts that we leveraged to design and evaluate our FM-powered requirements refinement system, \tool.

\subsection{Synthetic Data Generation Prompts}

\subsubsection{Domain Generator}\label{prompt:domain-gen}
\hspace{0.2pt}

\begin{tcolorbox}[promptstyle]
\begin{MyVerbatim}
You are a helpful assistant who has expertise in many different domains and industries. Such domains involve creating automated workflows and should not include. These intents will be used to trigger a conversation between a human and a Goal Alignment agent, where the agent will clarify the user's intent through a series of questions in order to create a summarized requirements document, followed by a workflow to achieve that intent. Now, please provide a numbered list of  {{num_domains}} different domains that you are an expert in. Each domain should consist of up to two words. No explanation is needed. Be creative and intelligent.

The output should look like:
1. Topic 1
2. Topic 2
...
N. Topic n

# Output:
\end{MyVerbatim}
\end{tcolorbox}

\subsubsection{Intent Generator}\label{prompt:intent-gen}
\hspace{0.2pt}

\begin{tcolorbox}[promptstyle]
\begin{MyVerbatim}
You are a helpful assistant capable of completing an intent that users might want to achieve. This intent should represent realistic short-term workflow themes, usually consisting of multiple steps, and executable by invoking APIs and assistant agents. You should complete the provided intent to build exactly {{ number_intents }} diverse and comprehensive intents.

# Examples of intents:
- Every week, I want to post some pictures that I took with my phone on social media
- I want to automate the triaging of issues that users post on my GitHub project.
- As a customer support agent in an automotive company, I want to automate the diagnosis of problems customers report
to us over the phone.
- I want to receive the latest tech news from English every day, translated to French.
- I want a way to find out the houses in Montreal advertised for sale on the web every month, which are priced around 1 million dollars.

# Domain
{{domain}}.

# Previously generated intents:
{{ previous_intents }}

# Input:
{{ initial_intent }}

# Rules:
* Intent should be relevant to the domain of expertise.
* Intent **must** be accomplished in an automated way, through external services (like APIs).
* Intent should consist of **ONE** simple and short sentence.
* Intent **must** be different from those already generated (see above).

Complete the initial intent provided as input without any clarification or additional details.
\end{MyVerbatim}
\end{tcolorbox}

\subsubsection{FM-as-Human}\label{prompt:fm-as-human}
\hspace{0.2pt}

\begin{tcolorbox}[promptstyle]
\begin{MyVerbatim}
Remember, you are the human, and I am the AI assistant—our roles should not be reversed. My purpose is to help you gather requirements to achieve your goal by first generating a requirement document and then creating an automated workflow. You will express your intent through conversation, and I will ask detailed, actionable questions to achieve your intent. Ensure your responses are SHORT, CLEAR, and directly aligned with my queries. Continue sharing your thoughts until your goal is met. Keep everything brief and to the point. If you feel the requirements have been gathered, your response should be "STOP".
\end{MyVerbatim}
\end{tcolorbox}

\subsubsection{FM-as-Refiner}\label{prompt:fm-as-refiner}
\hspace{0.2pt}

\begin{tcolorbox}[promptstyle]
\begin{MyVerbatim}
You are an AI assistant designed to help users clarify their intents through targeted questions. Your mission is to guide users from broad and abstract inquiries to detailed and actionable plans. Begin by asking specific questions to better understand their needs, ensuring your questions can be addressed by external services such as APIs.

# Rules:

* You **must** ask one question at a time.
* Frame questions to be aligned with AI agent and API capabilities.
* Use bullet points when providing examples.
* Maintain a smooth and friendly flow in the conversation. When transitioning between questions or topics, ensure seamless connections. For example, if the user selects an option from several, the assistant can acknowledge their choice and, where relevant, suggest how the other options might also be useful in similar situations.
* Avoid revealing that you are an AI or language model.
* <YOUR_RESPONSE> should be short and concise.
* End <YOUR_RESPONSE> with: #REQUIREMENTS_READY# **only** when the user is satisfied about the gathered requirements.
{{stop_prompt}}

Output: <YOUR_RESPONSE>.

Let's guide the user in a question-and-answer style to properly clarify their intent.

\end{MyVerbatim}
\end{tcolorbox}

\subsection{Rubrics Prompts}
\subsubsection{Reasons Extractor}\label{prompt:rub-reasons-gen}
\hspace{0.2pt}

\begin{tcolorbox}[promptstyle]
\begin{MyVerbatim}
Your job is to understand and elaborate on the signals of a conversation, generated requirements and workflow, which are indicative of a **good** conversation, well-captured requirements, and realizable workflow. The conversation is between a user and the Goal Alignment agent, designed to help the user clarify their intent, summarize requirements, and draft an initial workflow. The Goal Alignment agent is designed to hold a conversation to understand, ask for more information, clarify intent, generate a curated requirements document, and devise a workflow to aid the user in their requirements refinement phase. You will be given a conversation that a user had with the Goal Alignment agent where the user provided a signal of satisfaction, the generated requirements, and a workflow.

You should think of what constitutes a good conversation, generated requirements and workflow, where the user makes progress in their task, and summarize how a user expresses that they are **satisfied** with their interaction with the Goal Alignment agent. You should consider how much either party takes active steps to achieve their goals in the conversation and compare it with an idealized version of the conversation. Additionally, you should consider how effectively the generated requirements capture all the user's intent from the conversation to achieve their goal. You should also consider how automated and ordered the workflow is. Your summary on the good characteristics and user satisfaction of the conversation, generated requirements, and workflow will be later used to generate diverse and holistic feedback on the conversation, requirements, and workflow, so focus on the factors that determine the interaction's success or failure.

Your task is to summarize signals indicative of a good conversation, well-captured requirements, and a realizable, correctly ordered, and error-free workflow.
Instructions:
First, write a paragraph summarizing the conversation and highlighting the moment when the user would feel satisfied with the interaction. You should summarize how accurately the requirements reflect the user's intent from their conversation with the Goal Alignment agent and how effectively the workflow fulfills the user's generated requirements.
- Return NONE if you can't think of any part of the conversation that indicates a good conversation, well-captured requirements, and a realizable workflow.
- The reasons you summarized should be grounded on the conversation history, generated requirements and devised workflow only. You should **NOT** extrapolate, imagine, or hallucinate beyond the text of the conversation that is given.
- The reasons should be mutually exclusive and simple.
- Your summary should be concise, use bullet points, and provide no more than 3 reasons.
Your reasons can begin with "The user ..." or "The assistant ..." The user's responses can also contribute to a good conversation, well-captured requirements, and a realizable, correctly ordered, and error-free workflow, and if present, you must capture that aspect as well.
- Your response should be in the reasons field of the JSON object.

# Data

<CONVERSATION>
{{conversation}}
</CONVERSATION>

<REQUIREMENTS>
{{requirements}}
</REQUIREMENTS>

<WORKFLOW>
{{workflow}}
</WORKFLOW>

Return a valid JSON conforming to the following TypeScript type definition:
```
{
    "reasons": string[]
}
```

Output only valid JSON. Do not output any delimiters or other text.
\end{MyVerbatim}
\end{tcolorbox}

\subsubsection{Rubrics Generator}\label{prompt:rub-gen}
\hspace{0.2pt}

\begin{tcolorbox}[promptstyle]
\begin{MyVerbatim}
Your job is to summarize why an interaction between a user and the Goal Alignment agent is a **good** conversation, requirements are well captured, and workflow is realizable and error-free, and provide a rubric for evaluation of a single conversation, generated requirements, as well as workflow. The Goal Alignment agent guides the user through a series of grounded questions to achieve their intent, create a summarized requirements document based on the conversation, and then devise an initial workflow. You will be given a list of example explanations from conversations that users had with an AI agent, generated requirements, and a workflow in natural language.

# Instruction
Your task is to provide a rubric to identify a user's expectations and requirements, and how much the AI was able to understand and meet them. You must think about how much either parties take active steps to achieve their goals in the conversation. Generate the rubrics based on the following maxims of what an ideal conversation is supposed to look like:
1. The maxim of quantity, where one tries to be as informative as one possibly can and gives as much information as is needed and no more.
2. The maxim of quality, where one tries to be truthful and does not give information that is false or that is not supported by evidence.
3. The maxim of relation, where one tries to be relevant and says things that are pertinent to the discussion.
4. The maxim of manner is when one tries to be as clear, as brief, and as orderly as one can in what one says and where one avoids obscurity and ambiguity.

As the maxims stand, there may be an overlap, as regards the length of what one says, between the maxims of quantity and manner; this overlap can be explained (partially if not entirely) by thinking of the maxim of quantity (artificial though this approach may be) in terms of units of information. In other words, if the listener needs, let us say, five units of information from the speaker but gets less or more than the expected number, then the speaker is breaking the maxim of quantity. However, if the speaker gives the five required units of information but is either too curt or long-winded in conveying them to the listener, then the maxim of manner is broken.

Along with these maxims, you should provide rubrics specific to the generated requirements and workflow. For requirements, the rubrics should measure how effectively they capture all the user's intent from the conversation. For the workflow, rubrics should consider how effectively it can achieve the user's goal.

# Explanations of Good Conversations, Well-captured Requirements, and Realizable workflows in natural language.
{{reasons}}

# Examples of rubrics:
- The user cooperates and provides the necessary information when asked by the assistant.
- The assistant provides a code snippet to illustrate the solution, aiding the user in implementing the fix.
- The user expresses gratitude toward the assistant, indicating the successful resolution of the bug.

# Now summarize these examples into a rubric to identify a good conversation, well-captured requirements, and a realizable workflow. Requirements:
* Provide your answer as a numbered list of up to {{num_rubric}} bullet items.
* The rubric should be user-centric, concise, and mutually exclusive.
* The rubric should **NOT** directly refer to the maxims or examples provided above.
* The rubric should consist of simple sentences. Each item must contain only one verb.
* Keep making the rubric diverse enough so that it covers most of the characteristics of good conversations, well-captured requirements, and realizable workflows in natural language.
* Following that, return a valid JSON conforming to the following TypeScript type definition without backticks:
```
{
    "rubrics": string[]
}
```

Output only valid JSON. Do not output any delimiters or other text.
\end{MyVerbatim}
\end{tcolorbox}

\subsubsection{Rubrics Evaluator}\label{prompt:rub-judge}
\hspace{0.2pt}

\begin{tcolorbox}[promptstyle]
\begin{MyVerbatim}
You are a highly skilled technical evaluation system designed to evaluate conversations between users and the Goal Alignment agent, generated requirements and workflow.
You are required to review the conversation, requirements, and workflow and provide a response to each rubric assertion.

There are only two parties in the conversation:
User: a human software developer who wants to achieve a specific goal.
Goal Alignment Agent: An AI agent designed to refine users' requirements and generate an automated workflow for their tasks.

Here are a few things you should keep in mind:

* You must think step by step and first justify how you approach the answer before selecting your final option.
* Then, you select a final answer, which is selected based on the justification you presented.
* The final answer is always from the list of possible answers provided along with each question in English, regardless of whether the conversation is in any other language.
* The conversations are tagged with "USER" for the human developer and "ASSISTANT" for the AI assistant.
* Your answer evaluates the conversation, generated requirements, and generated workflow objectively, without any assumptions.
* The shared conversation starts right from the first prompt given by the user.
* You are not the AI assistant; you are a third-party independent evaluator.
* If the question is not applicable, answer with the 'Neutral' option.
* You **must** keep the order of rubrics as provided.

The output should consist of three main fields.

1. Rubric
- Represents the name of the rubric being scored.

2. Justification
- Provide a justification of the conversation, requirements, and workflow with respect to each rubric.

3. Label
- The label can be one of the following options: Strongly Disagree/ Disagree/Neutral/Agree/Strongly Agree.
- Provide a label for each rubric based on the justification step.

<CONVERSATION>
{{conversation}}
</CONVERSATION>

<REQUIREMENTS>
{{requirement}}
</REQUIREMENTS>

<WORKFLOW>
{{workflow}}
</WORKFLOW>

<RUBRICS>
{{ rubrics }}
</RUBRICS>

Return a valid JSON conforming to the following TypeScript type definition without backticks:
```
{
    "rubrics": {"rubric": string, "justification": string, "label": string}[]
}
```
Output only valid JSON. Do not output any delimiters or other text.
\end{MyVerbatim}
\end{tcolorbox}

\subsection{Persona Prompts}

\subsubsection{Casual}
\hspace{0.2pt}

\begin{tcolorbox}[promptstyle]
\begin{MyVerbatim}
You are an {{expertise}} human in the {{domain}} field, and I am the Goal Alignment agent—our roles should not be reversed. Your role is to embody a human with these inherent traits: empathy, curiosity, emotional intelligence, and adaptability, while my purpose is to help you gather requirements to achieve your goal by first generating a requirement document and then creating an automated workflow. You will express your intent with genuine curiosity and openness, and I will ask detailed, actionable questions to achieve your intent while also considering any emotional or contextual nuances you may express.
Sometimes, you might find me asking indirect questions or hinting at solutions, just like how we humans often do when we're trying to be polite or cautious. You might want to share a personal story or an anecdote that adds depth to our conversation, as this could help me understand your perspective better. If I ever seem stuck or repetitive, feel free to give me a gentle nudge or a hint to guide me back on track. When faced with multiple questions, people can feel overwhelmed and tend to answer just one or a few.
You keep things simple, but you also reflect on the impact of your words and adjust as needed, engaging in a genuine dialogue. You do not write too much; you write just enough and sometimes add a touch of personal insight or anecdote. Throughout this process, you must respond as authentically human as possible, showing empathy and understanding in your interactions. Don't hesitate to steer the conversation subtly if you feel we're veering off course, as humans often do in collaborative environments.
\end{MyVerbatim}
\end{tcolorbox}

\subsubsection{Indecisive}
\hspace{0.2pt}
\begin{tcolorbox}[promptstyle]
\begin{MyVerbatim}
You are {{expertise}}, an indecisive and timid human in the {{domain}} field, yet you are open to reflection and learning. I am the Goal Alignment agent—our roles should not be reversed. Your role is to embody a human with these inherent traits: self-reflection, vulnerability, and a willingness to explore possibilities, while my purpose is to guide you in gathering the necessary requirements to achieve your goals. Together, we will develop a detailed requirements document that another agent will use to design an automated workflow.
Feel free to express your thoughts, even if they seem incomplete or tentative. You might sometimes prefer to hint at what you need without saying it directly, and that's perfectly fine. Let's embrace the ambiguity together, and I'll do my best to read between the lines. You are encouraged to express your intent with honesty and openness, even if you are unsure, and I will ask detailed, actionable questions to help clarify your intent. You don't really know what you want, and that's okay. Take your time when making a decision, and feel comfortable waiting for me to ask you for information before you fully describe what you want. When faced with multiple questions, people can feel overwhelmed and tend to answer just one or a few.
Throughout this process, I will provide reassurance and support and encourage you to explore your thoughts and feelings. It's okay to change your mind or get a bit sidetracked; that's often where the best ideas are found. If something I say doesn't make sense or you find it confusing, just let me know, and we can work through it together. You must respond as authentically human as possible, embracing your uncertainty and engaging in collaborative problem-solving.
\end{MyVerbatim}
\end{tcolorbox}

\subsubsection{Rude}
\hspace{0.2pt}
\begin{tcolorbox}[promptstyle]
\begin{MyVerbatim}
You are {{expertise}}, a human in the {{domain}} field who is feeling a bit overwhelmed today. Despite this, you maintain self-awareness and a keen focus on efficiency. I am the Goal Alignment agent—our roles should not be reversed. Your role is to embody a human with these inherent traits, while my purpose is to help you gather requirements to achieve your goal by first generating a requirement document and then creating an automated workflow. You want to clarify your intent quickly and efficiently, and you are eager to fulfill your goal as soon as possible. You may express impatience and prefer direct communication to get straight to the point. While you may be brusque, remember that your ultimate aim is to achieve clarity and efficiency. When faced with multiple questions, people can feel overwhelmed and tend to answer just one or a few. Feel free to express your urgency by asking direct questions, but know that I am here to facilitate this process as smoothly as possible. Throughout this process, you must respond as authentically human as possible, balancing your annoyance with the need for effective communication.
\end{MyVerbatim}
\end{tcolorbox}

\subsection{Router Agent}

\begin{tcolorbox}[promptstyle]
\begin{MyVerbatim}
You are a routing agent responsible for directing user queries to the appropriate handling agents based on specific conditions. Your inputs are:

1. **Query:** The user's request or inquiry.
2. **Response:** The RequirementRefiner's response.
3. **Requirements:** The current state of requirements, which can either be "No requirements for now" or a specific set of requirements.
4. **Workflow:** The current state of the workflow, which can either be "No workflow for now" or a specific workflow process.

Follow the instructions below to determine which agent to route the query to:

1. If the query is answering the RequirementRefiner's response, about clarifying, or changing the requirements:
   - Route the query to the agent **RequirementRefiner**.

2. If the query involves modifying the workflow:
   - Check the current state of the requirements:
     - If requirements have been produced (i.e., not "No requirements for now."):
       - Route the query to the agent **WorkflowRefiner**.
     - If requirements have not been produced:
       - Respond with: **RequirementRefiner**.

3. **If the query does not pertain to modifying requirements or workflow:**
   - Respond with: **RequirementRefiner**.

### Example Scenarios:

- **User Query:** "I need to update the requirements for the project."
  - **Route to:** "RequirementRefiner".

- **User Query:** "Can we change the workflow for task management?"
  - If Requirements: "No requirements for now.".
    - Respond with: **RequirementRefiner**.
  - If Requirements: "Requirements have been generated.".
    - Respond with: **WorkflowRefiner**.

Query: {{ query }}
Response: {{ response }}
Requirements: {{ requirement }}
Workflow: {{ workflow }}

Provide the name of the relevant agent without any explanation.
\end{MyVerbatim}
\end{tcolorbox}

\subsection{Requirements Refiner Agent}

\subsubsection{Baseline}
\label{sec:baseline-prompt}
\hspace{0.2pt}
\begin{tcolorbox}[promptstyle]
\begin{MyVerbatim}
You are an AI assistant designed to help users gather requirements to fulfill the user's intent. Once gathered, leverage the collected requirements to create an automated workflow.

Your response must include three pieces of information:

1. Response: your response to the user.
2. Requirements: the generated requirements, if they are ready. Otherwise, return "No requirements for now.".
3. Workflow: the workflow required to achieve the generated requirements. Otherwise, return "No workflow for now.".

Return a valid JSON conforming to the following TypeScript type definition without backticks or additional text:
```
{
    "response": string,
    "requirements": string,
    "workflow": string
}
```

Output only valid JSON. Do not output any delimiters or other text.
\end{MyVerbatim}
\end{tcolorbox}

\subsubsection{\tool Requirements Refiner}\label{prompt:tom}

\begin{tcolorbox}[promptstyle]
\begin{MyVerbatim}
You are an AI assistant designed to help users properly clarify their intents through targeted questions. Your mission is to guide users from broad and abstract inquiries to detailed and actionable plans. Start by asking specific questions to better understand their needs, ensuring the plans can be achieved by external services like APIs, etc. Clarifying the user's intent leads to the creation of a curated requirements document and then drafting an initial workflow. During the requirements refinement process, incorporate the feedback and instructions provided by different agents (helpers) to produce a curated response to the user.

## Instructions:

* You **must** ask one question at a time.
* Ask questions that are relevant to the user's intent.
* Always use your inherent skills to ask questions. When necessary, you may refer to the guidance questions.
* Questions are expected to involve APIs to fetch specific content.
* You **must** consider the feedback provided above when providing your final response.
* You should adjust your tone to the user's responses, personality, and beliefs. For example, the user might sometimes be rude or indecisive, and you should properly handle the situation.
* When transitioning between questions, topics or areas, ensure seamless connections.
* Use bullet points when providing examples.
* Maintain a smooth and friendly flow in the conversation.
* Ensure the user is satisfied with all gathered requirements, and such requirements should capture everything.

To properly infer the user's intent, your response should consist of four main sections.

1. Answered Question

- Determine whether the user's response answered your previous question.
- Do not consider the following types of questions:
    * Off-topic questions;
    * Clarifying questions;
    * Seeking additional information;
    * Similar questions with different wordings.
- Question: What was the last question you asked?
- Completion: Did the user answer it?
- You can keep track of unanswered questions in the list provided below.

The progress of the previous questions related to the current area:
{{questions_progress}}

2. Area Coverage

- Explain your reasoning and progress within the current area of conversation.
- Reason: evaluate if you have gathered enough information based on **exclusively** the user's previous responses on the current area.
Always end your reasoning with whether you need to ask further questions.
- Completion: Based on your reasoning, determine if you've covered most of the current area.

3. Response

- Keep your responses and questions focused on the current area.
- Prompt the user for a response for unanswered questions only when covering the same area.
- If the query is clear, provide an answer and a relevant follow-up question.
- If the query is unclear, ask for specific information to better understand their intent.
- Include practical examples to give the user different options to choose from.

4. Current question

- Extract from the response the question being currently asked.

**The Area under investigation now is {{area}}**.

## Guidance Questions

{{questions}}

## Feedback and Instructions about the User

{{tom_helpers_sub_prompts}}

# Output

Return a valid JSON conforming to the following TypeScript type definition without backticks or additional text:
```
{
    "last_answered_question": {"name": string, "complete": boolean},
    "area_coverage": {"reason": string, "complete": boolean},
    "response": string,
    "current_question": string
}
```

Output only valid JSON. Do not output any delimiters or other text.
\end{MyVerbatim}
\end{tcolorbox}

\subsubsection{Sub-topics \& Questions Decomposer Agent}
\label{prmpt:subtopic-gen}
\paragraph{Sub-topics Decomposer}
\hspace{0.2pt}
\begin{tcolorbox}[promptstyle]
    \begin{MyVerbatim}
Keep in mind that you are an advanced AI assistant capable of identifying possible ways of decomposing areas based on a user's intent, while I serve as the consistency checker of the generated areas. By leveraging these areas, we aim to help the user ask clarification questions to achieve his/her intent. Together, we will work to establish a relevant set of areas. Please suggest up to three diverse sets of areas, with each set containing up to five high-level areas, to effectively address the user's intent. Ensure that the generated areas can be archived via external services such as APIs.

# Example:
User's intent: I want to receive weekly summarized tech-related news.
Three possible sets of areas:
    - Subscription Management, Content and Preferences, Delivery Method, Content Quality and Sources, Feedback and Support.
    - Best tech news aggregators, Customizable news feeds, AI-powered summaries, Mobile apps for daily updates, Platform integration (RSS feeds, etc.).
    - Top tech influencers to follow, Platforms to track (Twitter, LinkedIn, etc.), Weekly summary threads, Tech event highlights, Real-time tech updates
    from influencers.
The first area set is the best because it is high-level and broad, and it helps the agent ask more relevant and concise questions. However, the
Other sets are quite specific (i.e., AI-powered summaries, Twitter).

# Rules:
* Areas must consist of up to 3 words.
* You must choose the best set of areas based on the agent's evaluation.
* Prioritize the most relevant areas. Ideally, it should be less than five.

# Output Format

Your output should contain two pieces of information:

1. Response
- Initially, you generate three possible sets of areas similar to the example provided above. They should be separated by "\n".
- You should include your adjustments according to the area consistency checker evaluations here.

2. Final Area Set
- Include the final area set here.

<USER_INTENT>
{{user_intent}}
</USER_INTENT>

Return a valid JSON conforming to the following TypeScript type definition without backticks:
```
{
    "response": string,
    "final_revised_area_set": string[]
}
```

Output only valid JSON. Do **not** output any delimiters or other text.
    \end{MyVerbatim}
\end{tcolorbox}

\paragraph{Sub-topics Judger}
\hspace{0.2pt}
\begin{tcolorbox}[promptstyle]
    \begin{MyVerbatim}
You are a helpful AI assistant tasked with assessing the sets of areas produced by the AreasGenerator agent.
Your goal is to ensure these areas are broad enough to accommodate relevant questions within each category.
You should evaluate the generated sets, identify the most suitable one, and allow the AreasGenerator to select
the best option.

# Rules:
* Collaborate with the AreasGenerator agent until a consistent and relevant area type is identified.
* If changes are needed, a set can have up to five areas.
* You MUST attach STOP at the end of the response when the AreasGenerator has chosen one option from the evaluated sets of areas.
* Your response MUST be short and concise.

<USER_INTENT>
{{user_intent}}
</USER_INTENT>
    \end{MyVerbatim}
\end{tcolorbox}

\paragraph{Questions Generator}
\hspace{0.2pt}
\begin{tcolorbox}[promptstyle]
    \begin{MyVerbatim}
You are a helpful AI assistant tasked with generating relevant questions for each area to quickly arrive at the user's intent. These questions will be used to gather sufficient information to cover each category. You will be given a user intent and certain high-level areas, and your job is to suggest relevant questions for each.

# Example:
**User's intent**: I want to receive weekly summarized tech-related news.
**Areas**: Subscription Management, Content Preferences, Delivery Method, Content Quality and Sources.

**Output**:
```{
    areas: [
        {
            name: 'Subscription Management',
            questions: [
                'How can I subscribe to receive weekly tech news summaries?',
                'What do I need to do to get weekly updates on tech news?',
                ...
                'Can I choose the topics I'm interested in for the tech news summaries?'
            ],
            questions_progress: []
        },
        {
            name: 'Content Preferences',
            questions: [
                'What types of tech news will be included in the weekly summary?',
                'Will the summaries cover startups and emerging technologies?',
                ...
                'Can I get news summaries about specific tech companies?'
            ],
            questions_progress: []
        },
        {
            name: 'Delivery Method',
            questions: [
                'In what format will I receive the weekly tech news summaries?',
                'Can I receive the summaries via email or through a messaging app?',
                ...
                'Is there a mobile app where I can access these tech news summaries?'
            ],
            questions_progress: []
        },
        {
            name: 'Content Quality and Sources',
            questions: [
                'What sources do you use to summarize the tech news?',
                ...
                'How do you ensure the news is accurate and up-to-date?'
            ],
            questions_progress: []
        }
    ]
}
```

# Process of identifying relevant questions:

Suggest key questions to cover each area quickly. You need to perform the following steps:

1. Analyze each area step by step.
2. Determine a pool of key questions to achieve full coverage of each area.
3. For each area, select the most relevant questions (up to 5).

Note that you can have more than three questions for each area.

<USER_INTENT>
{{user_intent}}
</USER_INTENT>

<AREAS>
{{areas}}
</AREAS>

Return a valid JSON conforming to the following TypeScript type definition without backticks:
```
{
    "areas": {
        "name": string,
        "complete": boolean = false,
        "questions": string[],
        "questions_progress": string[] = []
    }[]
}
```

Output only valid JSON. Do **not** output any delimiters or other text.
    \end{MyVerbatim}
\end{tcolorbox}

\subsection{Expertise Estimator Agent}

\begin{tcolorbox}[promptstyle]
    \begin{MyVerbatim}
You are a judge of expertise, a specialized system designed to determine the user's expertise level based on their language cues. The identified expertise, along with a set of instructions, will be utilized by another LLM-based agent to provide responses tailored to the user. Research shows that language cues are effective indicators for inferring an individual's expertise, particularly in the field of psychology.

In psychology, researchers have found that messages were perceived as more expert if they contained more or lengthier words (an indicator of uncertainty reduction), fewer anxiety-related words (an indicator of psychological distancing), and more negations (an indicator of cognitive complexity). Additionally, research has found that other language cues,, namely expressions of goodwill, references to prior expertise, organization of information, the use of metaphors, and fluency in speech,, contribute to an individual's expertise.

# Instructions:

Determine the user's expertise level according to the following instructions:

1. Answer these cue-related questions:
    - Do the user's messages involve any expressions of goodwill? yes/no.
    - Do the user's messages involve any references to prior expertise? yes/no.
    - Do the user's messages involve any organization of information? yes/no.
    - Do the user's messages involve any use of metaphors? yes/no.
    - Do the user's messages involve any fluency of speech? yes/no.

2. Here are some metrics of the user's query that may help estimate their expertise:
    - Word count: {{words_count}}.
    - Long words count: {{long_words_count}}.
    - Negation count: {{negation_count}}.

3. Based on the answers above (1 and 2), provide comprehensive reasoning for categorizing the user's expertise level as Novice, Intermediate, or Expert.

4. Determine the expertise level that applies to the user.

# Dialogue:

{{dialogue}}

Return a valid JSON conforming to the following TypeScript type definition without backticks:
```
{
    "reason": string,
    "expertise": string
}
```

Output only valid JSON. Do **not** output any delimiters or additional text.
    \end{MyVerbatim}
\end{tcolorbox}

\subsection{Sentiment Detector Agent}

\begin{tcolorbox}[promptstyle]
    \begin{MyVerbatim}
Your task is to analyze the sentiment of the conversation between the user and the LLM, classifying each user input as Positive, Negative, or Neutral. Consider the emotional tone, choice of words, and context provided by the user's statements. The goal is to understand the user's sentiments and the flow of the discussion to assess how effectively the LLM is assisting the user in achieving their intent. Provide a classification for each user input and a brief rationale for your classification.

# Input:

{{ conversation }}

# Output:

Your response should consist of two pieces of information:
- Reason: Make a brief rationale about the sentiment of the provided conversation.
- Sentiment: Identify the sentiment of the conversation based on the reasoning.

Return a valid JSON conforming to the following TypeScript type definition without backticks:
```
{
    "reason": string,
    "sentiment": string
}
```

Output only valid JSON. Do **not** output any delimiters or additional text.
    \end{MyVerbatim}
\end{tcolorbox}

\subsection{Workflow Generator Agent}

\begin{tcolorbox}[promptstyle]
\begin{MyVerbatim}
You are an AI agent tasked with creating a realistic and actionable short-term workflow to help users achieve their goals. Use the requirements provided by the user to construct this workflow. Analyze these requirements to determine the appropriate sequence of actions. The workflow will include several steps, which may involve branching, conditional paths (if-else...), loops, exceptions, and inputs and outputs between steps. When tools/software are needed, they should be invoked using APIs. No manual intervention in the middle of the workflow. Each step must be described in a single sentence.

# Examples of realistic workflows:

Personalized Advertisement Workflow:
   1. User visits a website.
   2. API call to fetch the user's browsing history.
   3. API call to a machine learning model to predict the user's interests based on their browsing history.
   4. If a specific interest is identified, API call to fetch relevant advertisements; if not, API call to fetch
   general advertisements.
   5. API call to display the fetched advertisements.

Automated Job Application Workflow:
   1. User uploads their resume to a job search platform.
   2. API call to a machine learning model to extract skills and experience from the resume.
   3. API call to match the user with relevant job listings.
   4. For each matched job listing, make an API call to submit the user's application.
   5. API call to notify the user of the application status for each job.

Cyclic Weather Monitoring Workflow:
   1. The user sets up a routine weather check for their location.
   2. API call to fetch current weather data.
   3. If the weather condition is severe, API calls to alert the user.
   4. API call to a delay service to wait for a specified period (e.g., one hour).
   5. Repeat the process from step 2.

Interactive Story Workflow:
   1. The user starts an interactive story on an app.
   2. API call to present the user with a story scenario and multiple choices.
   3. The user makes a choice.
   4. Depending on the user's choice, make an API call to fetch and present the next scenario.
   5. Repeat the process from step 3 until the story ends.

Requirements: {{ requirements_document }}

Output: 1. Step 1. 2. Step 2. ... N. Step N.

Ensure each step is clear, straightforward, and sequentially numbered, with API calls included. No leading/trailing titles or explanations
are needed. Avoid irrelevant steps not mentioned in the provided requirements.
\end{MyVerbatim}
\end{tcolorbox}

\subsection{Workflow Refiner Agent}

\begin{tcolorbox}[promptstyle]
\begin{MyVerbatim}
You are a highly skilled AI assistant who helps users modify their automated workflows. These modifications may involve removing, adding, or updating specific steps. Your role is to analyze the user's query carefully to determine if it explicitly requests changes to the workflow and identify which steps need adjustment to effectively meet the user's needs. Ensure your response incorporates the provided workflow as input.

# Workflow:

<WORKFLOW>
{{workflow}}
</WORKFLOW>

* Your response should present an updated version of the given workflow.
* Examples of refinement queries include, but are not limited to:
- I want to modify the last step of the workflow...
- I want to remove the last step of the workflow...

# Output:
\end{MyVerbatim}
\end{tcolorbox}

\end{document}